\documentclass[12pt]{iopart} 
\usepackage[utf8]{inputenc}
\usepackage[english]{babel}

\usepackage{graphicx}

\expandafter\let\csname equation*\endcsname\relax
\expandafter\let\csname endequation*\endcsname\relax
\usepackage{amsmath}

\usepackage[colorlinks=true,allcolors=blue]{hyperref}
\usepackage{cite}

\begin{document}

\title{Burstiness in activity-driven networks and the epidemic threshold}

\author{Marco Mancastroppa,$^{1,2}$ Alessandro Vezzani,$^{3,1}$ \\Miguel A. Mu\~noz$^{4,1}$ and Raffaella Burioni $^{1,2}$}
\address{$^1$ Dipartimento di Scienze Matematiche, Fisiche e Informatiche, Università degli Studi di Parma, Parco Area delle Scienze, 7/A 43124 Parma, Italy}
\address{$^2$ INFN, Gruppo Collegato di Parma, Parco Area delle Scienze 7/A, 43124 Parma, Italy}
\address{$^3$ IMEM-CNR, Parco Area delle Scienze, 37/A, 43124 Parma, Italy}
\address{$^4$ Departamento de Electromagnetismo y F{\'\i}sica de la Materia e Instituto Carlos I
  de F{\'\i}sica Te\'orica y Computacional. Universidad de Granada.
  E-18071, Granada, Spain}
\eads{\mailto{marco.mancastroppa@unipr.it}, \mailto{alessandro.vezzani@unipr.it},  \mailto{mamunoz@onsager.ugr.es} and \mailto{raffaella.burioni@unipr.it}}

\begin{abstract}
We study the effect of heterogeneous temporal activations on epidemic spreading in temporal networks. We focus on the susceptible-infected-susceptible (SIS) model on activity-driven networks with burstiness.
By using an activity-based mean-field approach, we derive a closed analytical form for the epidemic threshold for arbitrary activity and inter-event time distributions. We show that, as expected, burstiness lowers the epidemic threshold while its effect on prevalence is twofold.  In low-infective systems burstiness raises the average infection probability, while it weakens epidemic spreading for high infectivity. Our results can help clarify the conflicting effects of burstiness reported in the literature. We also discuss the scaling properties at the transition, showing that they are not affected by burstiness. 
\end{abstract}

\vspace{2pc}

%\maketitle

\section{Introduction}\label{sez:intro}
The heterogeneous distribution of times between two consecutive actions -- often denoted as \textit{burstiness} -- is a typical signature of time-resolved records of human activities \cite{barabasi2005origin,karsai2018bursty}.  Mobile phone calls, social contacts and many other complex real-world dynamics exhibit patterns of enhanced activity within short time periods, followed by long lags of inactivity.  Actually, the frequency of events shows large variability and strong temporal fluctuations \cite{barabasi2005origin,goh2008burstiness,vazquez2006modeling,karsai2012universal} and the event dynamics strongly differs from a Poisson process, where the rate at which events occur is constant. In ``bursty'' processes, the probability distribution for the inter-event time between two successive actions is not an exponential as in Poisson processes but, typically, it features a fat tail, easily measured from large datasets \cite{barabasi2005origin,karsai2018bursty}.

Recent studies have addressed the problem of the origin of this highly heterogeneous behavior, arising from the decision-based queuing process that humans apply in performing tasks and allocating their priority \cite{karsai2012universal,malmgren2008poissonian}. Besides, a large amount of work has been devoted to clarify the effects of intermittent patterns on the temporal structures of interactions, as described by temporal networks \cite{holme2013temporal}. Bursty dynamics can indeed influence the structure of links on the local and on the global scale \cite{ubaldi2017burstiness,burioni2017asymptotic,moinet2015burstiness,moinet2016aging,colman2015memory}. More importantly, heterogenous temporal patterns in the evolution of time-varying networks can affect in a non-trivial way dynamical processes. Analytical arguments \cite{akbarpour2018diffusion,vazquez2007impact,iribarren2009impact} and numerical approaches \cite{karsai2011small,rocha2011simulated} have indeed shown that burstiness can significantly modify processes mediated by interactions, such as random walks, epidemics, information diffusion, consensus formation, percolation. Among these, epidemic spreading is one of the most representative and widely applicable example \cite{pastor2015epidemic}.  Analytical results on epidemics in temporal network have focused on the epidemic threshold \cite{valdano2015analytical,valdano2018epidemic}, in general pushed to lower values by bursty effects. In other works, burstiness is introduced on a static network as a non-Markovian effect in infection (or recovery) processes \cite{starnini2017equivalence,boguna2014simulating,liu2018burst,van2013non,cator2013susceptible}.

Interestingly, when comparing bursty with Poisson dynamics in epidemics, 
{conflicting observations have been reported. On the one hand,
numerical results} \cite{vazquez2007impact,rocha2011simulated,karsai2011small} and modeling techniques \cite{iribarren2009impact,rocha2013bursts,van2013non,jo2014analytically} provided strong evidence of a slowing down for the late time spreading in the presence of burstiness, while opposite  effects are reported in the early-time dynamics \cite{masuda2013predicting}. In \cite{vazquez2007impact} and \cite{karsai2011small}, burstiness is found to slow down spreading in early times, while other works observe the opposite effect \cite{rocha2013bursts,rocha2011simulated}.

To understand {such conflicting observations} and keep track at the same time of the temporal evolution of interactions, in this paper we focus on the Susceptible-Infected-Susceptible (SIS) process in the presence of burstiness on activity-driven networks \cite{perra2012activity,starnini2013topological,liu2014controlling}. In activity-driven networks,  the propensity to engage
an interaction is moved from the links to nodes ($N$), by assigning to each node $i$ its activity potential $a_i$ measuring the {typical number} of activations (link formations) per time performed by agent $i$. We then model the bursty behavior by a fat tailed distribution of inter-event times between two successive activations of the same node. Using an activity based mean-field approach, which is here exact in the large $N$ limit, we find a closed analytical form for the epidemic threshold for general activity and inter-event time distributions including, as a particular case, bursty dynamics. The analytical results are in excellent agreement with extensive numerical simulations. We show that, as expected, burstiness lowers the epidemic threshold, while its effect on prevalence is twofold: burstiness tends to raise the average stationary infection probability  in low infective systems, therefore strengthening the epidemic, while it weakens the epidemic in high infective systems, lowering the prevalence. This result can help to clarify the conflicting effects of burstiness reported in the literature. We also analyze in details the scaling behavior at the transition, showing that the critical exponents are {mean-field exponents of the directed percolation (DP) universality class} and that they are unchanged in the presence of burstiness.

The paper is organized as follows. In Section \ref{sez:model} we describe the activity-driven network with bursty dynamics and we define the epidemic process unfolding on top of it. In Section \ref{sez:threshold} we present the analytical approach and we derive an exact expression for the epidemic threshold, comparing the result with extensive numerical simulations. In Section \ref{sez:bursty_effects} we analyze the effects of burstiness on disease spreading, focusing on the epidemic threshold as well as on prevalence, and we discuss the scaling properties. Finally, Section \ref{sez:conclusions} contains a summary of our results and some perspectives of our work.

\section{Epidemics on activity-driven networks}\label{sez:model}
\subsection{Activity-driven temporal networks}

We introduce the \textit{activity-driven network} by associating to each node $i=1,\dots,N$ an activity $a_i$ drawn from a distribution $\rho(a_i)$. The activation rate of node $i$ is defined by its inter-event time distribution $\Psi_{a_i}(\tau)$ \cite{ubaldi2017burstiness,burioni2017asymptotic,moinet2015burstiness,moinet2016aging,colman2015memory}: this sets the statistic of  time intervals between consecutive activations of node $i$, so that the average inter-event time equals the inverse of the activity of node $i$, i.e. $\langle \tau \rangle_i=a_i^{-1}$. The network evolves  according to a Gillespie-like algorithm \cite{gillespie1976general,boguna2014simulating,masuda2018gillespie}. We initialize the system at $t=0$, drawing from $\Psi_{a_i}(t)$ the first activation time, $t_i$, of each node. Then the agent $i$ with the lowest $t_i$ becomes active and create $m$ links connecting to 
$m$ randomly chosen nodes. We fix the next activation time $t_i$ for node $i$ as $t_i + \tau$
with $\tau$ drawn from $\Psi_{a_i}(\tau)$. Finally all  links are removed and the process is iterated, activating the node with smallest activation time $t_i$.

Empirical measures on real datasets provide a strong evidence of heterogeneous activity patterns, with $ \rho(a_i) $ being a broad, heterogeneous function with large fluctuations \cite{ubaldi2016asymptotic,perra2012activity,karsai2014time}. We model such a heterogeneity in the activity distribution as a power law: 
\begin{equation}
\rho(a_i) = \nu  a_m^{\nu} a_i^{-(\nu+1)}
\label{activity}
\end{equation}
for $a_i>a_m$ and $\rho(a_i)=0$ for $a_i<a_m$, where $a_m$ is a lower cut-off.  However, as we show in the next sections, our results hold for any $\rho(a_i)$ and therefore for any activity-driven network.

The inter-event time distribution sets the dynamics of activation for each site. For example,
an exponential $\Psi_{a_i}(\tau) =a_i e^{-a_i \tau}$ leads to a Markovian Poisson process as in standard continuous time activation dynamics.
When the inter-event time distribution of a single node is heterogeneous, with trains of activation events with short time separation, followed by long inactivity, 
we take into account the bursty behavior choosing a power-law (Pareto) distribution of inter-event times with a cut-off $\xi_i$: 
\begin{equation}
\Psi_{a_i}(\tau) = \alpha \xi_i^{\alpha} \tau^{-(\alpha+1)}
\label{Burst}
\end{equation}
for $\tau> \xi_i$ and $\Psi_{a_i}(\tau) =0$ for $\tau<\xi_i$.  {More specifically; for $\alpha>1$, we define $\xi_i = \frac{\alpha-1}{\alpha a_i}$ so that $\langle \tau \rangle_i=a_i^{-1}$, and the distribution of $\xi_i$ is $\Phi(\xi_i) \sim \rho(\frac{\alpha-1}{\alpha \xi_i}) \xi_i^{-2}$. Doing so, we can effectively compare systems with the same average inter-event time, fixed by the inverse of the node activity, but with different burstiness distributions $\Psi_{a_i}(\tau)$.  In the case $0<\alpha<1$, $\langle \tau \rangle_i$ diverges and the activity is not properly defined.  We can nevertheless introduce the inter-event time distribution according to Eq. (\ref{Burst}). As in \cite{burioni2017asymptotic,ubaldi2017burstiness,moinet2016aging} we set $\xi_i=a_i^{-1}$, with the cut off $\xi_i$ distributed among the nodes according to $\Phi(\xi_i) = \xi_i^{-2} \rho(\xi_i^{-1})$. However, in this case $a_i$ does not correspond to the activity of the node, i.e. the inverse of average activation time, and non-trivial aging effects may be present \cite{moinet2016aging}. In particular, for the SIS model, since node activation becomes always more unlikely as time increases, we will show that epidemics do not spread in the system.}

\subsection{The SIS epidemic model on activity-driven networks}

We consider the epidemic Susceptible-Infected-Susceptible (SIS) model. Each node can be infected ($I$) or susceptible ($S$-healthy) and
the dynamics is based on two elementary processes: contagion, when a susceptible individual, coming in contact with an infected person, has a probability $\lambda$ to contract the infection $S+I \xrightarrow[]{\lambda} 2I$; spontaneous recovering, when an infected turns susceptible with a rate $\mu$: $I \xrightarrow[]{\mu} S$.  The SIS model on activity-driven networks is defined by a Gillespie-like algorithm as follows \cite{perra2012activity,liu2014controlling}): \begin{enumerate} 
\item {Up to a relaxation time, $t_0$, the network evolves in the absorbing state where all nodes are susceptible, so that the node activation dynamics relaxes to its equilibrium.  
\item At time $t = t_0$ we start the epidemics setting the initial condition, for which we divide the population into a configuration of susceptible (S) and infected (I).}    
\item \label{item:3} Agent $i$ with the lowest activation time $t_i$ activates. Infected nodes at time $t$ are recovered at $t_i$ with probability $1-e^{-\mu (t_i-t)}$, i.e. the recovery is a Poisson process. Then we set $t=t_i$.  
\item The active node $i$ generates $m$ links with other randomly-chosen nodes. If the nodes involved in the contacts (links) are both infected ($I$) or susceptible ($S$) nothing happens, otherwise a contagion can occur with probability $\lambda$: $S+I \xrightarrow[]{\lambda} 2I$. Here we consider $m = 1$, with no loss of generality. 
\item A new inter-event time for the active node $i$ is drawn from $\Psi_{a_i}(\tau)$, hence the new activation time $t_i$ would be $t_i + \tau$. Then all its links are removed and the process is iterated from point \ref{item:3}.  
\end{enumerate}

As well known, the SIS model features a phase transition in the class of directed percolation (DP) between an absorbing phase, where the system evolves towards a state with the whole population susceptible (healthy), and an endemic phase, with  the epidemics spreading persistently in the population \cite{pastor2015epidemic,henkel2008non}. We consider a control parameter $r$  (e.g. $r=\frac{\lambda}{\mu}$ on static networks, $r=\frac{\lambda \langle a \rangle}{\mu}$ on activity-driven networks with Poisson dynamics \cite{perra2012activity,liu2014controlling}), so that the phase transition occurs at a critical value  $r_C$, the \textit{epidemic threshold}: if $ r<r_C $ the system is in the absorbing phase, while if $r> r_C$ the system is in the endemic phase. This is the quantity that we will  investigate in the next section.

\section{The epidemic threshold}\label{sez:threshold}
\subsection{Analytical results}
We consider the SIS model on activity-driven networks with arbitrary $\rho(a)$ and $\Psi_a(\tau)$.  If the average inter-event time $\langle \tau \rangle $ is finite (i.e. $\alpha>1$ considering the power-law distribution of Eq. \eqref{Burst}), nodes can be identified according to their activity, so we divide the population into classes with the same $a_i=a$ and study the probability that a node belonging to a class $a$ is infected at time $t$. This corresponds to an activity-based mean-field approach \cite{perra2012activity,tizzani2018epidemic}, which turns out to be exact for the present system given that connections are continuously reshuffled at each time step, destroying local correlations.

{At each contact event occurring at time $t$, we consider the following three quantities to fully characterize the system state:}\\
- $Q_a(t)$: the probability  for a node in the class $a$, that activates at time $t$, to be infected right before $t$;\\
-  $P_a(t)$: the probability for a non-active node, with activity $a$, to be infected at $t$;\\
- $Z_a(t,t')$: the probability for a node with activity $a$ to be infected at time $t$, knowing that its last activation occurred at time $t'<t$.

The Master Equation for the evolution of $Z_a(t,t')$ can be written as:
\begin{equation}
\partial_t Z_a(t,t')=-\mu Z_a(t,t') + \lambda (1-Z_a(t,t')) \int_0^{\infty} da' \rho(a') a' Q_{a'}(t)
\label{eq:ME_Z}
\end{equation}

where first term accounts for the recovery process with rate $\mu$, while the second term describes (for $\alpha>1$) the contagion process; actually, this is the product of the probability to be susceptible at time $t$, $(1-Z_a(t,t'))$, times the probability to be infected by an active node, which, for $\alpha>1$, can be expressed as
$\lambda \int_0^{\infty} da' \rho(a') a' Q_{a'}(t)$ \footnote{{ The probability for node $i$ to infect a node in the interval  $[t,t+dt]$ is $\lambda n_i(t)Q_i(t)/N$, where $n_i(t)$ is the number of activations of node $i$ in $dt$. For $\alpha>1$  we have $n_i(t)= dt/\langle \tau\rangle_i =a_i dt$, where $\langle \tau\rangle_i $ is the mean time between two activations.  Thus, the probability to be contacted by any infected node in $dt$ is $N^{-1} \sum_{i=1}^N a_i \, dt \, Q_i(t) = dt \int da' \rho(a') a' Q_{a'}(t)$.}}  Notice that this expression --together with the whole analytic approach-- does not apply for $0<\alpha<1$, since the average inter-event time diverges and $a_i dt$ is not the probability of activation of node $i$ in the interval $dt$. We observe that in Eq. \eqref{eq:ME_Z} we only include passive contagion events because, by definition of $Z_a(t,t')$, the node does not activate between times $t'$ and $t$.  

Eq. \eqref{eq:ME_Z} needs to be solved using the initial condition at time $t'$:
\begin{equation}
Z_a(t',t')=Q_a(t')+\lambda [1-Q_a(t')] \int_0^{\infty} da' \rho(a') P_{a'}(t')
\label{eq:CI_Z}
\end{equation}
where, in analogy with the equation above, the first term corresponds to the case where the active node with  activity $a$ is already infected at the contact time $t'$; the second term is the probability $(1-Q_a(t'))$ that the node is susceptible before $t'$, then it connects to an infected node with probability  $\int_0^{\infty} da' \rho(a') P_{a'}(t')$, and it is infected with probability $\lambda$.

{
Eq. \eqref{eq:ME_Z} with initial condition \eqref{eq:CI_Z} can be solved in the stationary condition, in the hypothesis that the system reaches an asymptotic steady state $(P_a^0,Q_a^0)$  with 
$(P_a(t),Q_a(t)) \xrightarrow[t \rightarrow \infty]{} (P_a^0,Q_a^0)$. In this regime, also the node activation dynamics needs to reach the steady state to avoid aging effect in the network evolution. For $\alpha> 1$ we expect this condition to be fulfilled if the relaxation time is much larger than the average activation time  ($t_0 \gg \langle a \rangle^{-1}$).
}
In the stationary regime, one can consider the averages on the activities as independent of time
$\int_0^{\infty} da \rho(a) a Q_{a}(t) \sim \int_0^{\infty} da \rho(a) a Q_{a}^0 = \overline{aQ}$ and
$ \int_0^{\infty}  da \, \rho(a) \, P_a(t)\sim \int_0^{\infty} da \rho(a) P_{a}^0 = \overline{P}$
to obtain:
\begin{equation}
Z_a(t,t')= \left[ Q_a(t') + \lambda \overline{P} (1-Q_a(t')) - \frac{\lambda \overline{aQ}}{\mu+\lambda \overline{aQ}} \right] e^{-(\mu+ \lambda \overline{aQ})(t-t')} + \frac{\lambda \overline{aQ}}{\mu + \lambda \overline{aQ}}
\label{eq:solZ}
\end{equation}

It is now possible to write explicitly $Q_a(t)$ and $P_a(t)$ as functions of $ Z_a(t, t') $
using the relations: \begin{eqnarray} Q_a(t) &=& \int_0^{\infty} Z_a(t,t-\tau') \Psi_a(\tau') d\tau' \nonumber \\ P_a(t) &=& \frac{1}{\int_0^{\infty} d\tau' \int_{\tau'}^{\infty} \Psi_a(\tau'') d\tau''} \int_0^{\infty} d\tau' Z_a(t,t-\tau') \int_{\tau'}^{\infty} d\tau'' \Psi_a(\tau''), \label{eq:QP} \end{eqnarray}
and plugging the solution \eqref{eq:solZ} into Eqs. \eqref{eq:QP}, one obtains:
\begin{eqnarray}
P_a(t)&=& \frac{\lambda \overline{aQ}}{\mu + \lambda \overline{aQ}} + a \left[ \lambda \overline{P} - \frac{\lambda \overline{aQ}}{\mu + \lambda \overline{aQ}} \right] \frac{1-L_a(\mu+ \lambda \overline{aQ})}{\mu + \lambda \overline{aQ}} \nonumber \\
&+& a [1-\lambda \overline{P}] \int_0^{\infty} d\tau' Q_a(t-\tau') e^{-(\mu+\lambda \overline{aQ})\tau'} \int_{\tau'}^{\infty} d\tau'' \Psi_a(\tau'') \label{eq:system} \\
Q_a(t)&=&\frac{\lambda \overline{aQ}}{\mu + \lambda \overline{aQ}}+ \left[ \lambda \overline{P} - \frac{\lambda \overline{aQ}}{\mu + \lambda \overline{aQ}} \right] L_a(\mu+ \lambda \overline{aQ}) \nonumber \\
&+& [1 - \lambda \overline{P}] \int_0^{\infty} d\tau'  Q_a(t-\tau') e^{-(\mu+\lambda \overline{aQ})\tau'} \Psi_a(\tau') \nonumber
\end{eqnarray}
where $L_a(\mu)=\int_0^{\infty} d\tau \, e^{-\mu \tau} \Psi_a(\tau)$ is the Laplace transform of $\Psi_a(\tau)$. In order to determine the epidemic threshold, one needs to study the stability of the absorbing state solution $(P_a(t),Q_a(t))=(0,0) \, \forall \, a$ of the Eqs. \eqref{eq:system}.

Solving Eqs. \eqref{eq:system} for $ t \rightarrow \infty $, two self-consistency equations for $\overline{P}$, $\overline{aQ}$ are obtained:
\begin{equation}
\begin{split}
\overline{P}&= h(\overline{P},\overline{aQ})= \frac{\lambda \overline{aQ}}{\mu + \lambda \overline{aQ}} + \int_0^{\infty} da \rho(a) \, a \frac{1-L_a(\mu+ \lambda \overline{aQ})}{(\mu+\lambda \overline{aQ})^2}  \frac{\lambda \mu \overline{P}}{1-(1-\lambda \overline{P})L_a(\mu+ \lambda \overline{aQ})}\\
\overline{aQ}&= f(\overline{P},\overline{aQ})= \int_0^{\infty} da \, \rho(a) \, a \frac{\left[ \frac{\lambda \overline{aQ}}{\mu + \lambda \overline{aQ}} + \left( \lambda \overline{P} - \frac{\lambda \overline{aQ}}{\mu + \lambda \overline{aQ}} \right) L_a(\mu+ \lambda \overline{aQ}) \right]}{1-(1-\lambda \overline{P})L_a(\mu+ \lambda \overline{aQ})}   
\end{split}
\label{eq:mappa}
\end{equation}
 (see \ref{sez:Conti} for a detailed derivation).
Eqs. \eqref{eq:mappa} can be considered as a two dimensional map: $(\overline{P}_{n+1}, \overline{aQ}_{n+1})={\cal M}(\overline{P}_{n}, \overline{aQ}_{n})$. The fixed point  $(\overline{P}_{n},\overline{aQ}_{n})=(0,0)$ corresponds to the solution $(P_a(t),Q_a(t))=(0,0) \, \forall \, a$ in Eq. \eqref{eq:system}. Then, the stability of the absorbing state can be studied by linearizing  ${\cal M}(\overline{P}_{n}, \overline{aQ}_{n})$ in $(\overline{P}_{n},\overline{aQ}_{n})=(0,0)$:
\begin{equation*}
\begin{bmatrix}
\overline{P}_{n+1}\\[6pt]
\overline{aQ}_{n+1}
\end{bmatrix}
\sim
J
\begin{bmatrix}
\overline{P}_{n}\\[6pt]
\overline{aQ}_{n}
\end{bmatrix}
=
\begin{bmatrix}
\frac{\lambda\langle a \rangle}{\mu} & \frac{\lambda}{\mu} \\[6pt]
\lambda \int_0^{\infty} da \rho(a) a \frac{L_a(\mu)}{1-L_a(\mu)} & \frac{\lambda \langle a \rangle}{\mu}
\end{bmatrix}
\begin{bmatrix}
\overline{P}_{n}\\[6pt]
\overline{aQ}_{n}
\end{bmatrix}
\end{equation*}
The absorbing state is stable if all the eigenvalues of $J$ are smaller than unity, i.e.:
\begin{equation}
\frac{\lambda \langle a \rangle}{\mu} + \lambda \sqrt{\frac{1}{\mu} \int_0^{\infty} da \rho(a) a \frac{L_a(\mu)}{1-L_a(\mu)}}<1.
\label{eq:disug_generale}
\end{equation}

Eq. \eqref{eq:disug_generale} provides an analytical form of the epidemic threshold for the SIS model on an activity-driven network ($m = 1$) with arbitrary $\rho(a)$ and $\Psi_a(\tau)$. As we will show, numerical simulations agree with the analytical estimate, suggesting that the activity based mean-field approach and the stationary hypothesis are exact for this model.

In particular, for an exponential distribution of inter-event times $\Psi_a(\tau) = a e^{-a \tau}$ and an arbitrary activity distribution $\rho(a)$, one can get the epidemic threshold from Eq.\eqref{eq:disug_generale}: 
\begin{equation}
\frac{\lambda\langle a \rangle}{\mu} = \frac{\langle a \rangle}{\langle a \rangle + \sqrt{\langle a^2 \rangle}}
\label{eq:pois}
\end{equation}
which is the known result for the SIS model on activity-driven networks with standard non-bursty dynamics \cite{perra2012activity}.
 
In the general case, the parameters $\lambda$, $\mu$ and $\langle a \rangle$ are coupled in a non-trivial way, since  Eq. \eqref{eq:disug_generale}
depends on the whole functional form of the activity and inter-event time distributions, so that the control parameter of the phase transition is not easily identified.
Hereafter, we will use the adimensional control parameter $r= \frac{\langle a \rangle}{\mu}$. Notice that, unlike the exponential case, the critical value depends in a non-trivial way on $\lambda$.

For a bursty distribution of inter-event times $\Psi_a(\tau) \sim \tau^{-(\alpha+1)}$  and the power law activity distribution $\rho(a) \sim a^{-(\nu+1)}$ defined by Eqs. \eqref{Burst} and \eqref{activity} respectively, Eq. \eqref{eq:disug_generale} becomes:
\begin{equation}
 \lambda r + \lambda \sqrt{\nu \, \left( \frac{r(\nu-1)}{\nu} \right)^{\nu} \int_{\frac{r (\nu-1)}{\nu}}^{\infty} dx \frac{1} {x^{\nu}} \, \frac{\alpha \left( \frac{\alpha-1}{\alpha x} \right)^{\alpha} \Gamma \left(-\alpha, \frac{\alpha -1}{\alpha x} \right)}{1-\alpha \left( \frac{\alpha-1}{\alpha x} \right)^{\alpha} \Gamma \left( -\alpha, \frac{\alpha -1}{\alpha x} \right)} } <1
\label{eq:disug_bursty2}
\end{equation}
where $\Gamma(s,x)=\int_x^{\infty} t^{s-1}e^{-t} dt$ is the upper incomplete Gamma function. In this case, the epidemic threshold, $r_C$, can be obtained by solving numerically Eq.\eqref{eq:disug_bursty2} for $r$.

\subsection{Numerical simulations}
We verify with numerical simulations the analytic result \eqref{eq:disug_generale}, checking the validity of our hypotheses. Henceforth, we fix $\lambda=1$, with no loss of generality. { We first let the network evolve for a time $t_0$ so that network dynamics is relaxed to the equilibrium of the activation dynamics and it has no memory of the initial condition at $t=0$ (when all sites are considered to be active). For $\alpha>1$ this means that $t_0$ needs to be much larger than the average activation time $t_0 \gg \langle a \rangle^{-1}$.}

The epidemic model evolves following the steps described in Section \ref{sez:model}. The epidemic threshold is numerically computed with the lifetime (or lifespan) method \cite{boguna2013nature,mata2015lifespan}. The lifetime $\tau_{life}$ is the average time that the system takes to reach the absorbing state when infecting at $t_0$ a single node of the system, discarding in the average  all the endemic realizations which remain infected for an infinite time. In simulations, we consider as endemic a realization whose coverage (number of distinct, ever infected nodes) reaches $N/2$. Observe that fluctuations can cause finite-time recovering even in the endemic phase so that the lifetime as a function of $r$ displays a peak at the critical point $r=r_C$ {behaving like an effective susceptibility as discussed in \cite{mata2015lifespan}. Also, it has been shown that the most effective initial condition to obtain a good estimate of the threshold on an inhomogeneous network is to infect at $t_0$ only the most connected site\cite{boguna2013nature}; this is the strategy we adopt here, infecting the most active one.}
\begin{figure}
\centering
\includegraphics[width=0.49\textwidth]{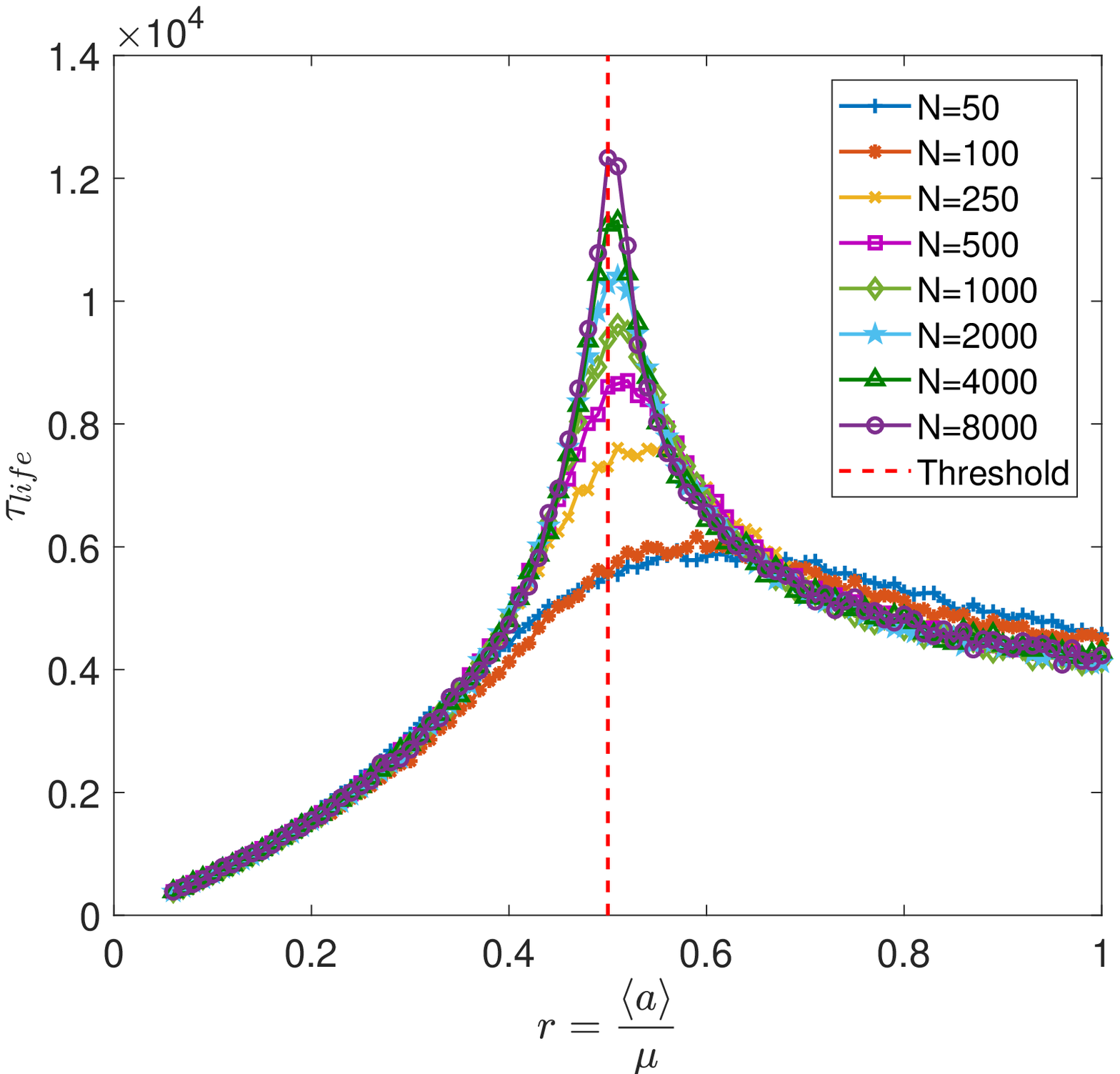}
\includegraphics[width=0.49\textwidth]{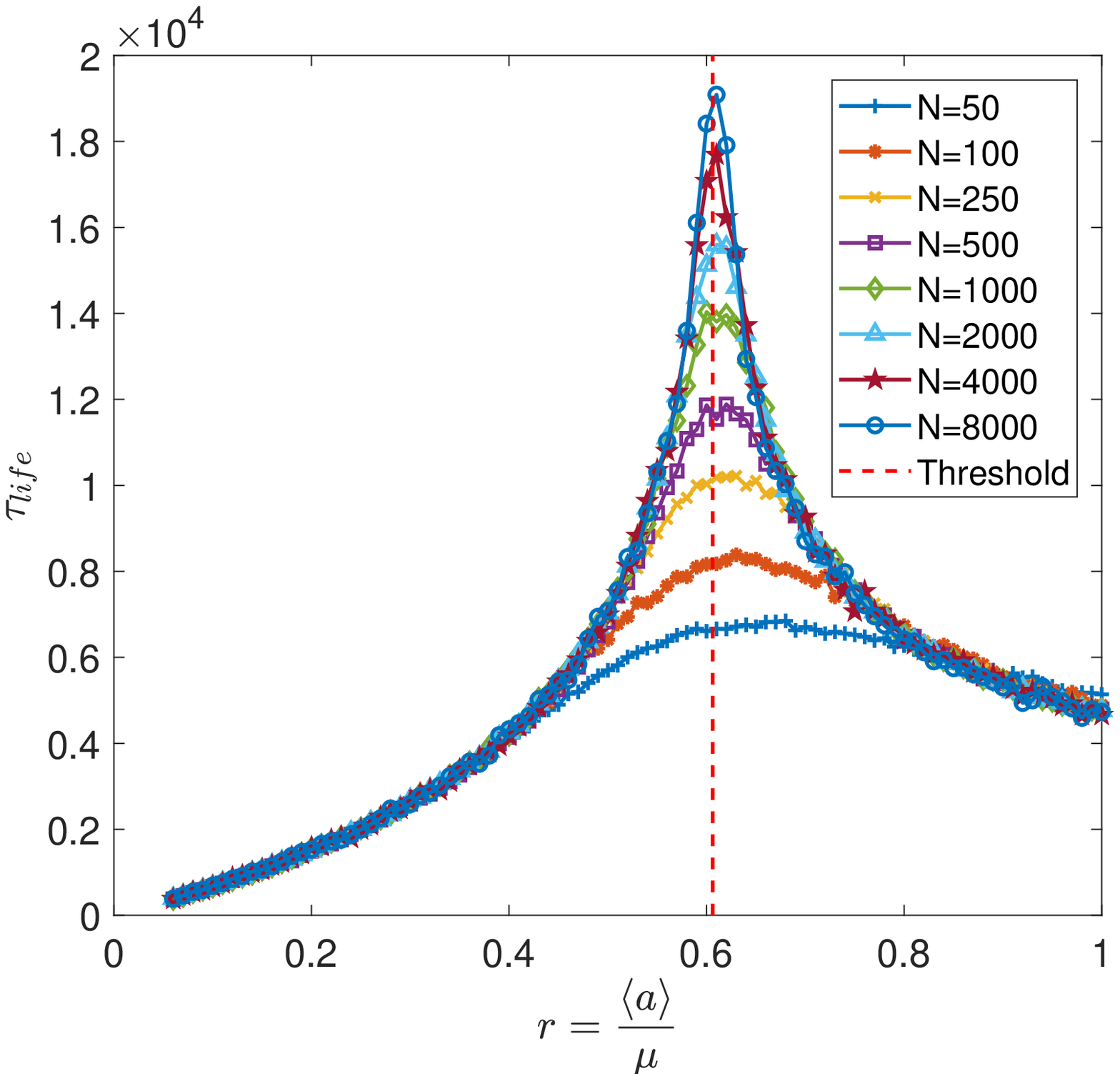}
\caption{Averaged lifetime $\tau_{life}$ as a function of the control parameter $r= \langle a \rangle/\mu$. Each value of $\tau_{life}$ is averaged over several simulations so that its uncertainty is less then $2 \%$. Different realizations of the network for $N \in [50,8000]$ are plotted. The red dashed line identifies the analytical epidemic threshold {as derived from Eq. \eqref{eq:disug_generale}}. In the left panel, we consider $\rho(a) \sim \delta(a-a_m)$, $a_m=1.65 \, 10^{-4}$ and $\Psi_a(\tau) \sim e^{-a \tau}$ (Poisson process), $\lambda=1$. In the right panel we set $\rho(a) \sim \delta(a-a_m)$, $a_m=1.65 \, 10^{-4}$ and $\Psi_a(\tau) \sim \tau^{-(\alpha+1)}$ (bursty dynamics), $\alpha=5$, $\lambda=1$.}
\label{fig:Numerico}
\end{figure}
Figure \ref{fig:Numerico} shows the lifetime as a function of the control parameter for different network sizes $N$, setting all nodes with same activity (i.e. $ \rho (a) = \delta (a-a_m) $).
Left and right panels consider an exponential $\Psi_a(\tau) \sim e^{-a \tau}$ and a power-law $\Psi_a(\tau) \sim \tau^{-(\alpha+1)}$ distribution, respectively. In both cases, for sufficiently large $N$, the numerical data agree very well with the analytical prediction of the threshold.
\begin{figure}
\centering
\includegraphics[width=0.5\textwidth]{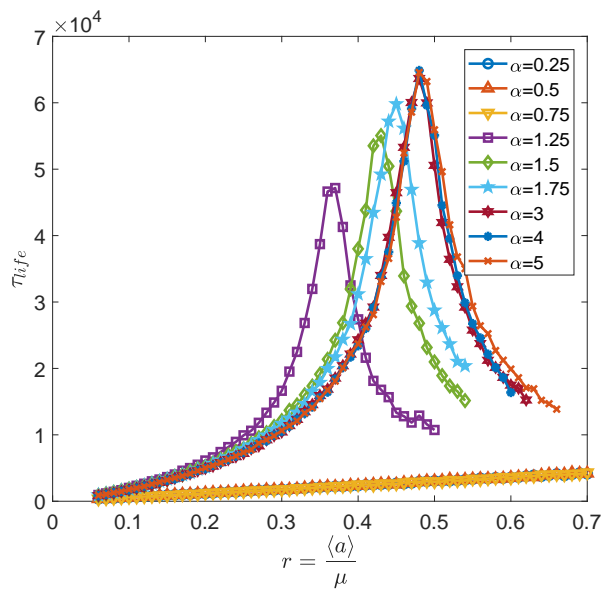}
\caption{Lifetime $\tau_{life}$ as a function of the control parameter $r= \langle a \rangle/\mu$ for an activity-driven network with $N=8000$, $\lambda=1$, $\rho(a) \sim a^{-(\nu+1)}$, $\nu=2.5$, lower cut-off $a_m=10^{-4}$ and $\Psi_a(\tau) \sim \tau^{-(\alpha+1)}$. Each value of $\tau_{life}$ is averaged over several simulations so that its uncertainty is less then $2 \%$. Each curve corresponds to a different exponent $\alpha \in [0.25,5]$; {observe that the three curves for $\alpha<1$ are hardly distinguishable, and that an active/endemic phase does not exist in such cases.}}
\label{fig:Alpha_diversi}
\end{figure}
Figure \ref{fig:Alpha_diversi} shows the lifetime as a function of $r$ for a network where $\rho(a) \sim a^{-(\nu+1)}$ and $\Psi_a(\tau) \sim \tau^{-(\alpha+1)}$ are power-law distributed according to Eqs. \eqref{activity} and \eqref{Burst} respectively. Several values of the exponent $\alpha$ have been considered. {When $0< \alpha<1$, as we discussed above, analytic results do not apply and we can introduce the inter-event time distribution $\Psi_{\xi_i}(\tau) \sim \xi_i^{\alpha}\tau^{-(\alpha+1)}$ with a broadly distributed  cut-off $\xi_i$ (as stated in Section \ref{sez:model}). Simulations reveal that the lifetime does not feature a peak since the epidemics always reaches the absorbing state. In particular, the average recovering time $\mu^{-1}$ is finite while node activation becomes always more unlikely as time increases, implying that, eventually, all nodes recover: an endemic or active phase does not exist in this case.}

\begin{figure}
\centering
\includegraphics[width=0.49\textwidth]{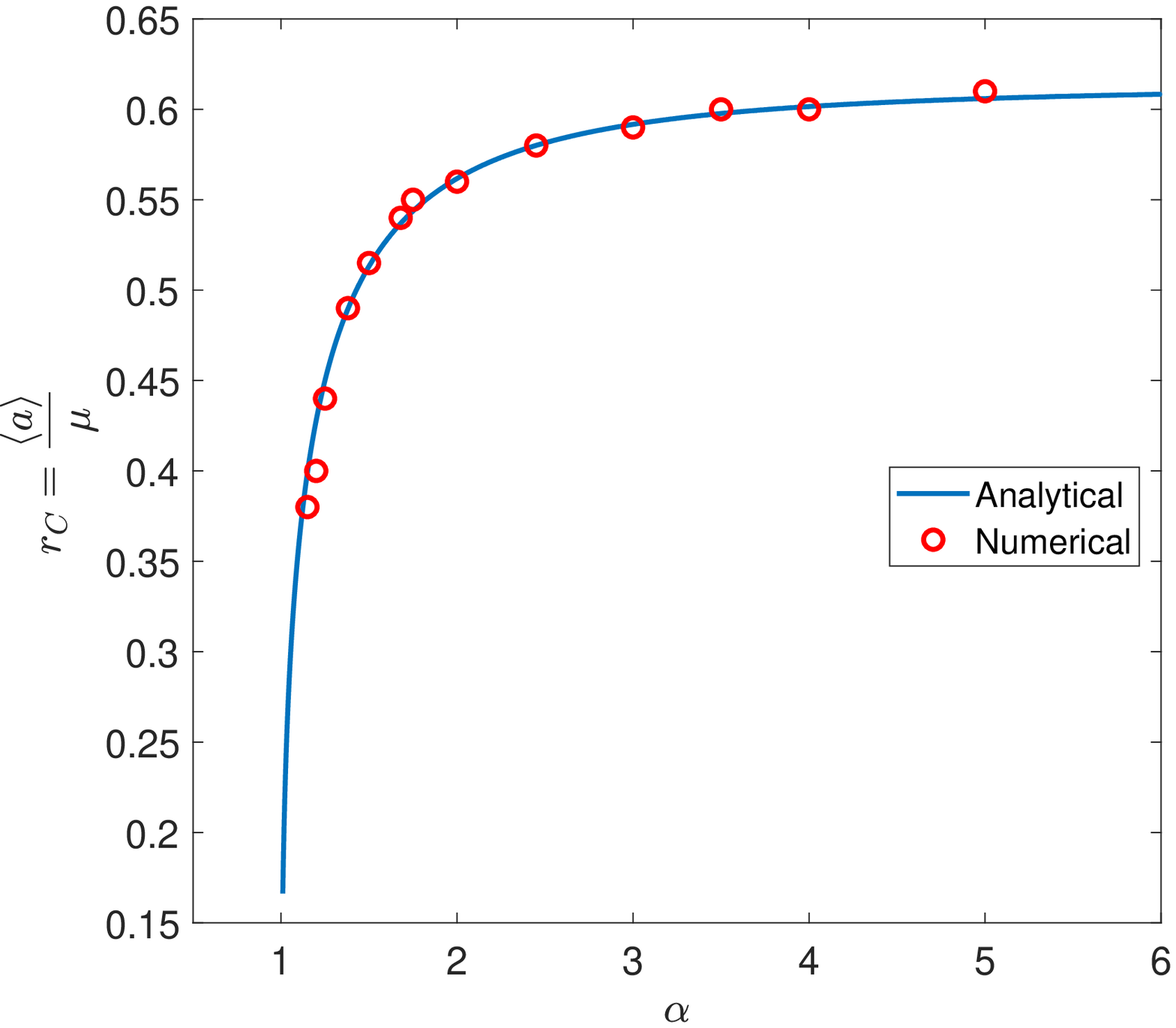}
\includegraphics[width=0.49\textwidth]{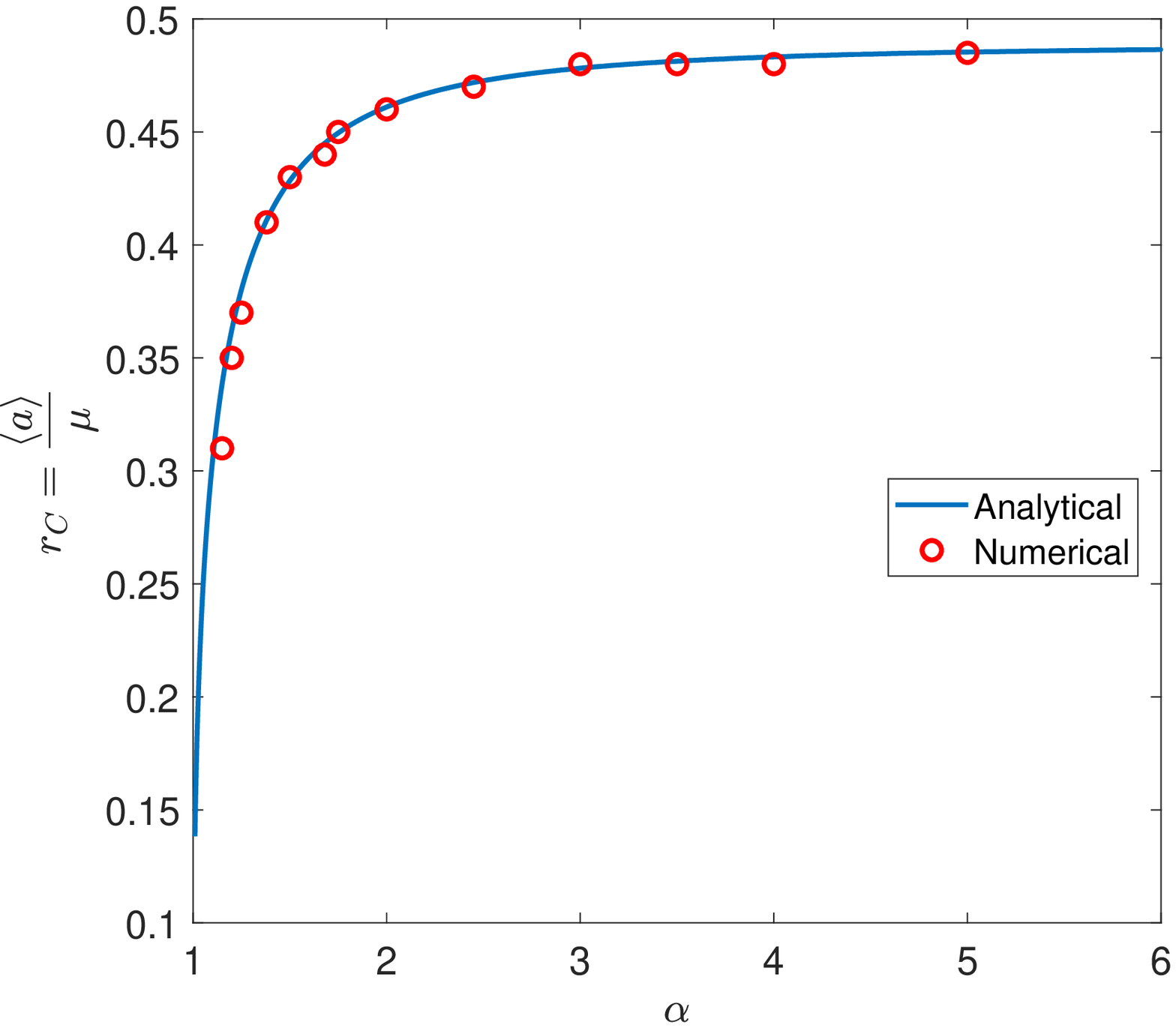}
\caption{Comparison between analytical (solid blue line) and numerical results (red dots) for the epidemic threshold. $r_C$ is plotted as a function of the exponent $\alpha$ of the inter-event time distribution $\Psi_a(\tau) \sim \tau^{-(\alpha+1)}$. In the left panel we set $\rho(a) \sim \delta(a-a_m)$, $a_m=1.65 \, 10^{-4}$, $\lambda=1$ and $N=4000$ in numerical simulations. In the right panel we consider $\rho(a) \sim a^{-(\nu+1)}$, $\nu=2.5$, lower cut-off $a_m=10^{-4}$, $\lambda=1$ and $N=8000$ in numerical simulations.}
\label{fig:Confronto_analitico_numerico}
\end{figure}

{Finally, in Figure \ref{fig:Confronto_analitico_numerico} we compare numerically and analytically obtained epidemic thresholds in the case of bursty dynamics for two different activity distributions $\rho(a)$, illustrating the excellent agreement between theory and numerics.} 

\section{Burstiness and epidemic spreading}\label{sez:bursty_effects}
\subsection{Epidemic thresholds for various distributions}

\begin{figure}
\centering
\includegraphics[width=0.5\textwidth]{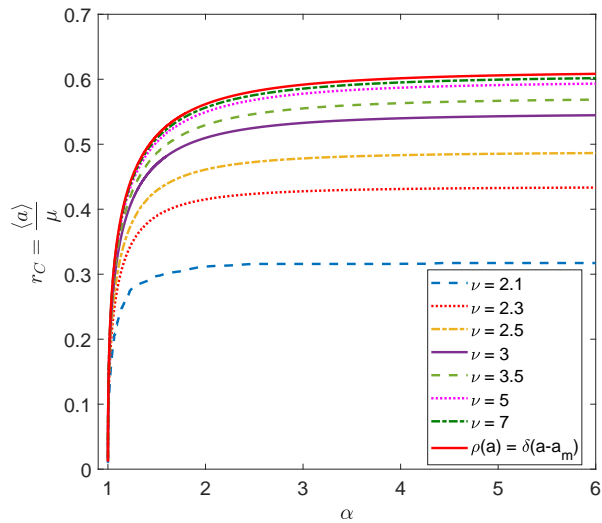}
\caption{Analytical epidemic threshold $r_C$ as a function of the exponent $\alpha$ of the inter-event time distribution $\Psi_a(\tau) \sim \tau^{-(\alpha+1)}$ for an activity-driven network with $\rho(a) \sim a^{-(\nu+1)}$. We plot $r_C(\alpha)$ for several values of $\nu \in [2.1,7]$, fixing the lower cut-off $a_m=10^{-4}$ and $\lambda=1$. {The solid red line corresponds to $\rho(a) \sim \delta(a-a_m)$, to which the other curves converge in the limit of large $\nu$.}}
\label{fig:Bursty_analitico}
\end{figure}

As explained above, for a generic distribution of inter-event times and activities, we use $r= \frac{\langle a \rangle}{\mu}$ as a control parameter for the phase transition, setting $\lambda=1$.  In particular, for a system with $\Psi_a(\tau)$ and $\rho(a)$ given by Eqs. \eqref{Burst} and \eqref{activity} respectively, the critical threshold $r_C$ is obtained solving Eq. \eqref{eq:disug_bursty2}; results are shown in Figure \ref{fig:Bursty_analitico}.  The behavior of $r_C$ is qualitatively the same for all $\nu$: it vanishes for $\alpha \rightarrow 1^+$ and it saturates to a constant depending on $\nu$ for $\alpha \rightarrow\infty$. { For $\alpha<1$, as explained above, the epidemics always ends up in the absorbing state.}

As expected, a broader activity distribution leads to a lower epidemic threshold and favors the propagation of the epidemics. In particular, for large $\nu$ the activity distribution tends to a $\delta$-function while in the limit $\nu \to 2$, i.e. when the activity fluctuations diverge, the epidemic threshold vanishes.

Let us now show that this is a general property for any inter-event distribution. Since $\int_0^\infty \tau \Psi_a(\tau) d\tau=a^{-1}$, $\Psi_a(\tau)$ tends to a $\delta$-function peaked in $\tau=0$ when $a\to \infty$. Hence, it is possible  to expand $e^{-\mu \tau}$ around $\tau=0$ when evaluating $L_a(\mu)$  for $a \to \infty$, in this way: $L_a(\mu) \xrightarrow[a \to \infty]{} 1$
and $1-L_a(\mu)=\int_0^\infty (1-e^{-\mu \tau})\Psi_a(\tau) d\tau \xrightarrow[a \to \infty]{} \int_0^\infty \mu \tau \Psi_a(\tau) d\tau= \mu/a$. Plugging these expression into Eq. \eqref{eq:disug_generale} we obtain that the integral over the activities diverges for $\nu<2$ due to the slow decay at large $a$.  Hence, Eq. \eqref{eq:disug_generale} cannot be satisfied 
for $\nu<2$ and the absorbing state is never stable. 
The behavior for $\nu \to 2$ and $\alpha\to 1 $  confirms that, keeping the average activity fixed, fluctuations and burstiness lower the epidemic threshold. 

\begin{figure}
\centering
\includegraphics[width=0.49\textwidth]{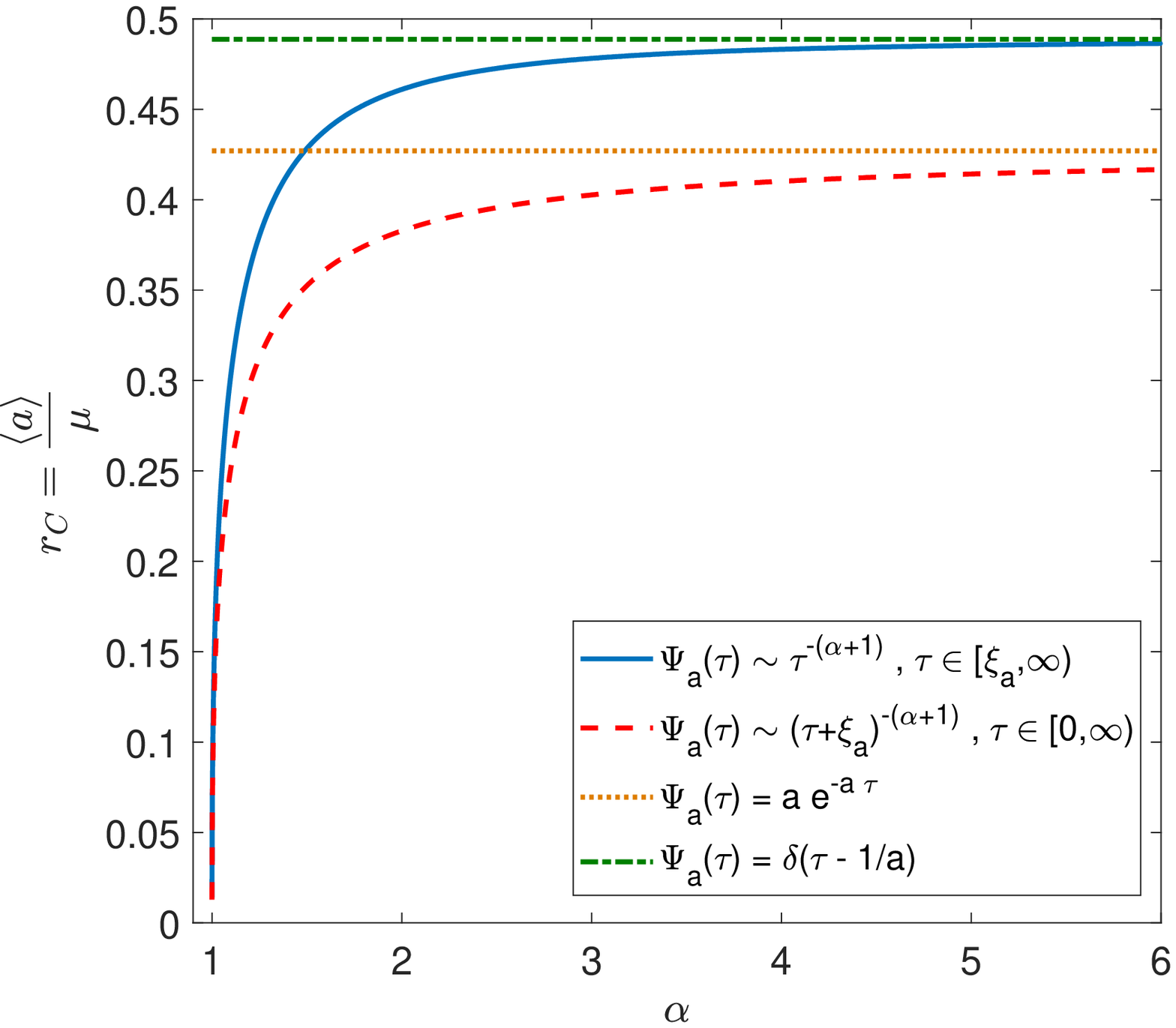}
\includegraphics[width=0.49\textwidth]{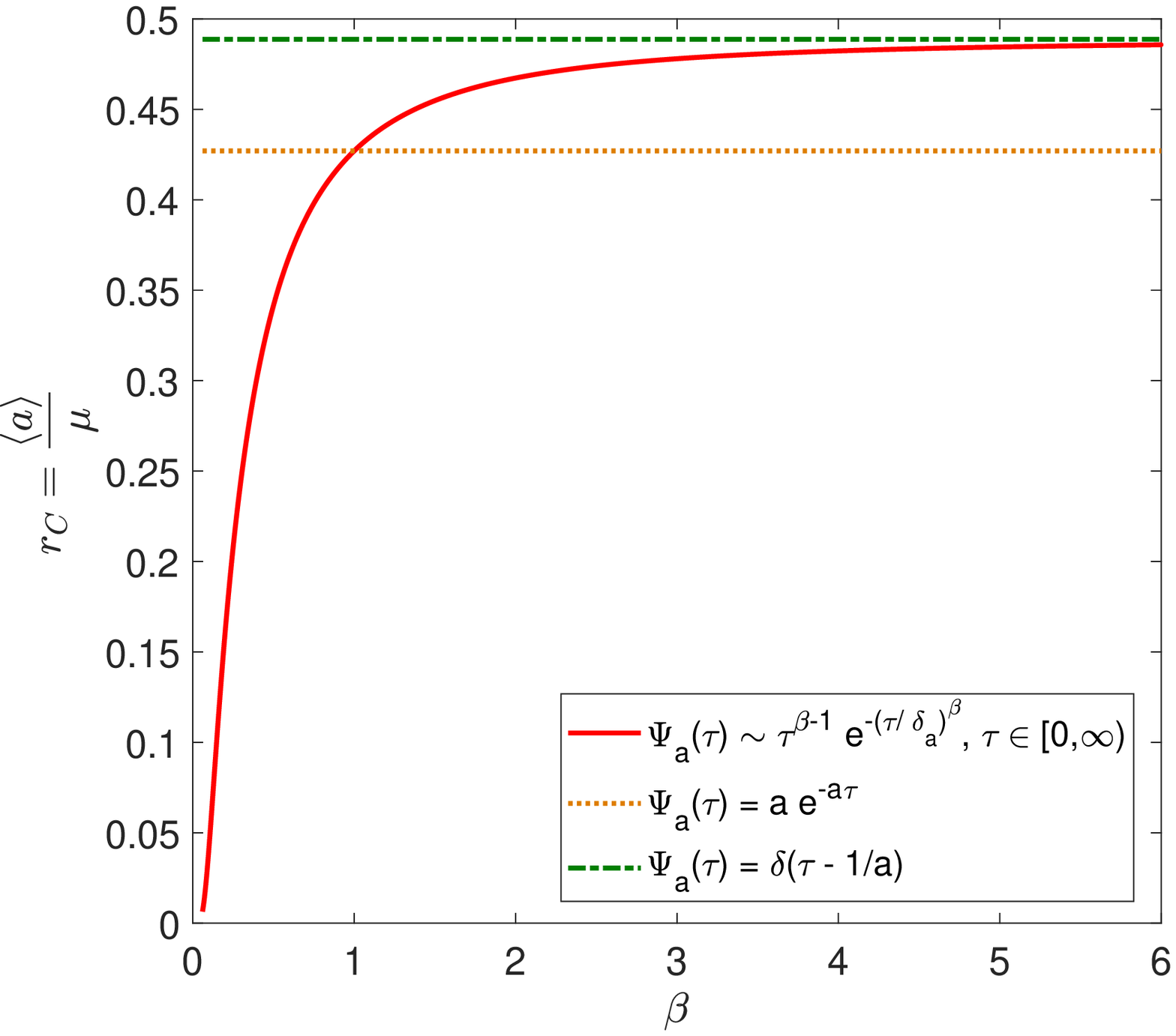}
\caption{{Epidemic threshold $r_C$ for different inter-event time distributions, fixing $\lambda=1$ and $\rho(a) \sim a^{-(\nu+1)}$ with $\nu=2.5$. In both panels: dotted orange line represents the exponential distribution $\Psi_a(\tau) = a e^{-a\tau}$; dash-dotted green line is the Dirac-delta function $\Psi_a(\tau) = \delta(\tau- 1/a)$. In the left panel the solid blue line is the Pareto distribution $\Psi_a(\tau)= \theta(\tau-\xi_a) \alpha {\xi_a}^{\alpha} \tau^{-(\alpha+1)}$ and the dashed red line is the a Lomax distribution $\Psi_a(\tau)= \theta(\tau) \alpha {\xi_a}^{\alpha} (\tau+\xi_a)^{-(\alpha+1)} $. In the right panel we plot the  Weibull distribution $\Psi_a(\tau)= \beta \delta^{-\beta} \tau^{\beta-1} e^{-(\tau/\delta)^{\beta}}$ with a solid red line.}}
\label{fig:Bursty_alternativa}
\end{figure}

{In Figure \ref{fig:Bursty_alternativa}, we show the epidemic threshold for a power-law activity distribution Eq. \eqref{activity} with $\nu=2.5$ and compare several inter-event time distributions $\Psi_a(\tau)$. In the left panel the continuous (blue) line is the epidemic threshold $r_C$ for inter-event times distributed according to the power law distribution with a cut-off in Eq. \eqref{Burst}, \emph{Pareto type I distribution}. The plot shows that for large $\alpha$ the threshold becomes even higher than in the exponential case, i.e. the dotted (orange) horizontal line. Indeed, for $\alpha \to \infty$, the power law distribution $\Psi_a(\tau)$ tends to a $\delta$-function and the node dynamics is periodic \cite{karsai2011small}, hence more regular than the Poisson dynamics (see the green dash-dotted line for comparison in Figure \ref{fig:Bursty_alternativa} - left panel). The dashed (red) curve in the left panel represents the epidemic threshold $r_C$ for the \emph{Lomax (or Pareto type II) distribution} $\Psi_a(\tau)= \alpha \xi_a^{\alpha} (\tau+\xi_a)^{-(\alpha+1)}$ with $\tau>0$: here $\xi_a=(\alpha-1)a^{-1}$ so that $\langle \tau \rangle_a=a^{-1}$, i.e. we introduce a different lower cut-off.  In this case, for $\alpha \to \infty$ we have that $\Psi_a(\tau)\to a e^{-a\tau }$, so in this limit the epidemic threshold tends to the value of the Poisson process (dashed orange line). The Lomax distribution is more heterogeneous than an exponential distribution, therefore the epidemic threshold is smaller for all $\alpha$. }

{Another interesting choice for the inter-event time --usually
considered in epidemic spreading--
 is the \emph{Weibull distribution} $\Psi_a(\tau)= \beta \delta_a^{-\beta} \tau^{\beta-1} e^{-(\tau/\delta_a)^{\beta}}$ with $\tau>0$ \cite{van2013non,liu2018burst}: here $\delta_a = (a \Gamma(1+1/\beta))^{-1}$ so that $\langle \tau \rangle_a = a^{-1}$. The Weibull distribution  interpolates between a broad distribution for $\beta \to 0^+$ (it tends to the Zipf distribution \cite{van2013non}) and a $\delta$-function for $\beta \to \infty$. Moreover, for $\beta=1$ it reduces to an exponential distribution. Notice that for any $\beta>0$ for large $\tau$ it decays faster than any power-law.
In the right panel of Figure \ref{fig:Bursty_alternativa} we plot $r_C$ for the Weibull distribution with a solid (red) line.
The threshold vanishes when $\beta \to 0$, it increases with $\beta$
and for $\beta=1$, $r_C$ coincides with that of the exponential case (dotted orange line).
For $\beta>1$ the threshold increases and for $\beta \to \infty$ it converges to the threshold of a periodic dynamics (dash-dotted green line).}

{ In general, from Figure \ref{fig:Bursty_alternativa}, we see that the value of $r_C$ depends on the whole shape of the inter-event time distribution, displaying, however, some general features.  For broad distributions, i.e. small $\alpha$ and $\beta$, the threshold is small and it vanishes for $\alpha\to 1$ and $\beta \to 0$.  Then, the threshold increases with $\alpha$ and $\beta$ and it reaches a maximum value for a periodic dynamics i.e. if $\Psi_a(\tau)$ tends to a $\delta$-function.  From here on, we consider a power-law distribution of inter-event times $\Psi_a(\tau)$ as defined in Eq. \eqref{Burst}.}

\subsection{Effects of burstiness on prevalence} 
\begin{figure}
\centering
\includegraphics[width=0.49\textwidth]{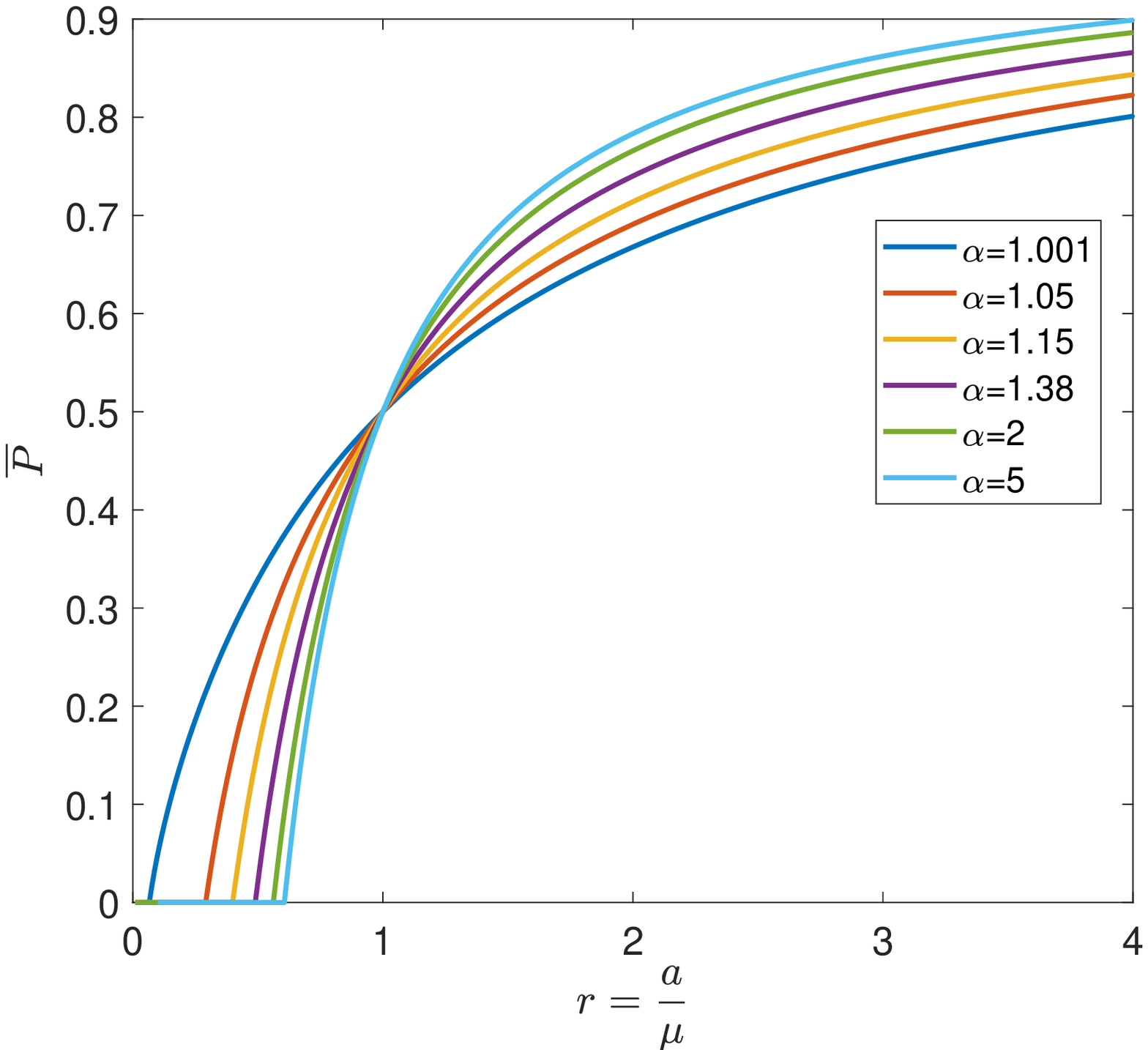}
\includegraphics[width=0.49\textwidth]{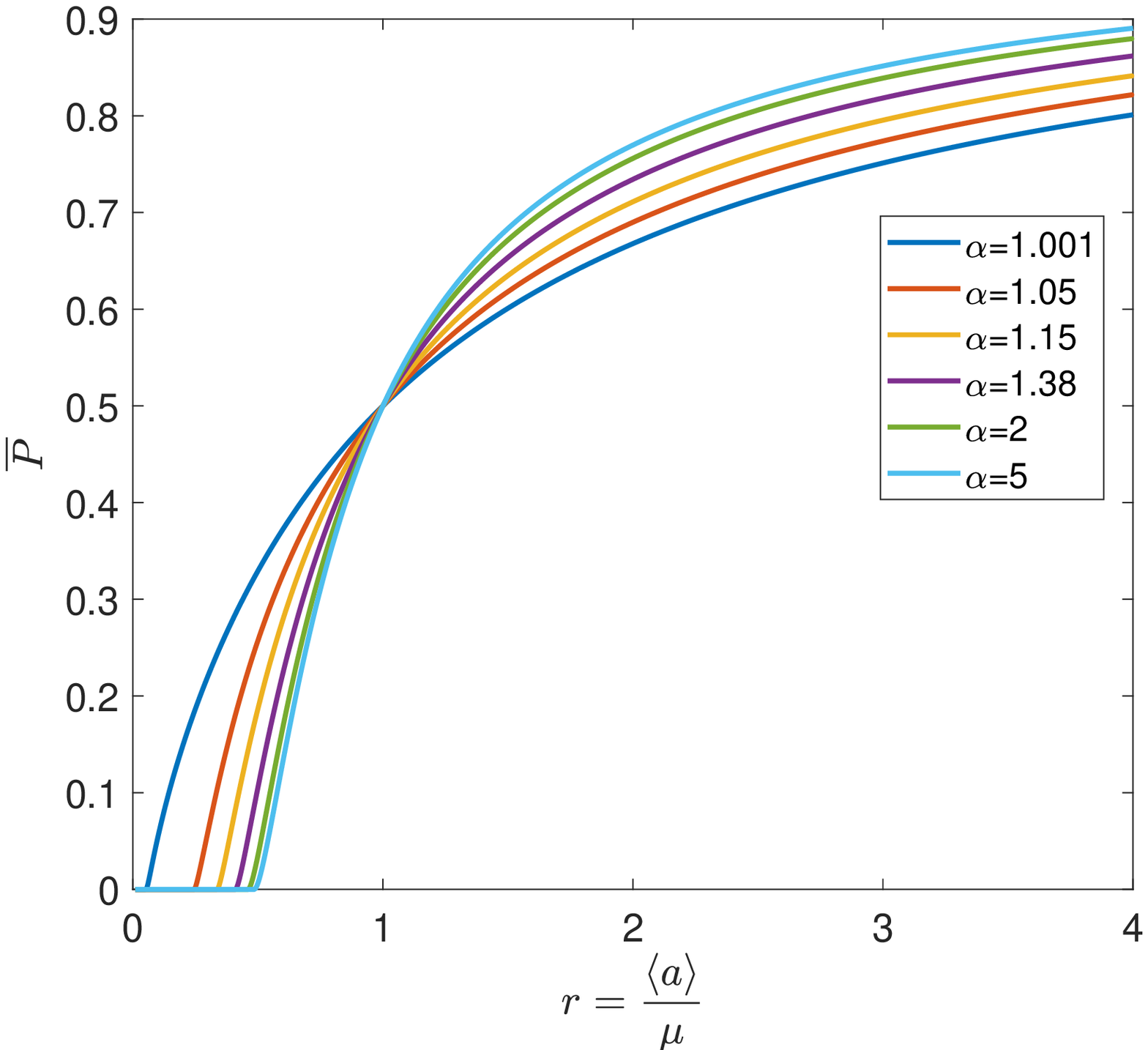}
\caption{Prevalence $\overline{P}$ as a function of the control parameter $r= \langle a \rangle/\mu$ considering an activity-driven network with bursty dynamics $\Psi_a(\tau) \sim \tau^{-(\alpha+1)}$. We plot $\overline{P}(r)$ for several values of $\alpha \in [1.001,5]$, fixing $\lambda=1$. In the left panel we set $\rho(a) \sim \delta(a-a_m)$ and the activity $a_m=10^{-4}$. In the right panel we consider $\rho(a) \sim a^{-(\nu+1)}$ with $\nu=2.5$ and lower cut-off $a_m=10^{-4}$.}
\label{fig:Prevalence}
\end{figure}

Within our analytic approach, beside the epidemic threshold value, a numerical solution of Eq. \eqref{eq:mappa} provides us with the value of $\overline{P}$ in the stationary state.
 This probability, called \textit{prevalence}, is the order parameter of the phase transition: it is zero below the epidemic threshold ($r<r_C$) and is different from zero above it ($r>r_C$).
In Figure \ref{fig:Prevalence} we plot the prevalence as a function of the control parameter $r = \frac{\langle a \rangle}{\mu}$  for different values of the burstiness exponent $\alpha$, when $\rho(a) \sim \delta(a-a_m)$ (left panel) and when $\rho(a) \sim a^{-(\nu+1)}$ as in Eq. \eqref{activity} (right panel).  

Plots confirm that burstiness lowers the epidemic threshold as discussed above. However, they also illustrate that the global effect of  burstiness ($\alpha \rightarrow 1$) is twofold: it raises the prevalence in low infective systems ($r <1$) strengthening the epidemic, while it weakens the epidemic in high infective systems ($r > 1$), lowering the prevalence. This behavior helps clarifying the conflicting effects of burstiness reported in the literature.  At $\lambda=1$ (see Fig. \ref{fig:Prevalence}), the switch from a regime where the burstiness strengthens the epidemics to the regime where epidemics is weakened occurs at $r=1$ and $ \overline{P}=1/2 $: in \ref{sez:prevalence} we show that, regardless the shape of $\Psi_a(\tau)$ and $\rho(a)$, this property holds when $ r=1/\lambda$. One can argue that the twofold effect of burstiness on prevalence is purely due to fluctuations. In this perspective, in \ref{sez:fluctuation} we show that the same behavior can be induced by activity fluctuation in systems without burstiness.

\subsection{Critical scaling behavior}
\begin{figure}
\centering
\includegraphics[width=0.49\textwidth]{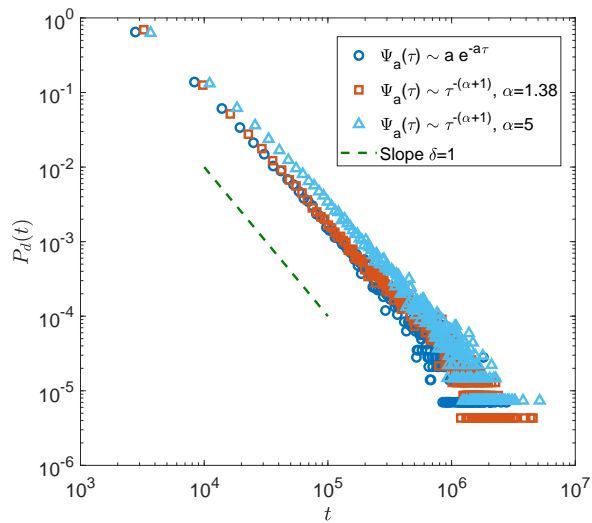}
\caption{The probability $P_d (t)$ as a function of time for a system at criticality.
We consider $\rho(a) \sim \delta(a-a_m)$, $\lambda=1$, $a_m=10^{-4}$ and several inter-event time distributions: $\Psi_a(\tau) \sim e^{-a\tau}$ (Poisson dynamics) and  $N=16000$, $\Psi_a(\tau) \sim \tau^{-(\alpha+1)}$ (bursty dynamics) with $\alpha=1.38$ and $N=8000$, or $\alpha=5$ and $N=10000$; we fix $\mu$ so that $r=r_C$ (critical point).}
\label{fig:delta}
\end{figure}

{
The study of so-called spreading experiments -- in which a single seed of infection is placed into an otherwise quiescent network -- is a standard tool to locate the phase transition between the active and the absorbing state. In such experiments one usually computes,  after the initial seed is placed, the probability $P_d(t)$ that a run reaches the absorbing state in a time interval $[t, t+dt]$, and the total number of infected nodes $N(t)$ at a given time $t$. These quantities scale as power laws at criticality, with exponents $\delta$ and $\eta$, respectively, defined by $P_d (t) \sim t^{-\delta-1}$ and $N(t) \sim t^{\eta}$ \cite{munoz1999avalanches}. In particular, in the DP universality class the mean-field exponents take the values $\delta=1$ and $\eta=0$ \cite{munoz1999avalanches}.  These are typically observed in high dimensional lattices, dense networks and fully connected systems.  In Figure \ref{fig:delta}, for $\rho(a) \sim \delta(a-a_m)$, we verify that the exponent $\delta$ in our activity driven model model is consistent with the mean field value $\delta \simeq 1$ both for exponential and power-law inter-event time distributions. A similar result can also be obtained for the exponent $\eta \simeq 0$. }

In finite size systems, the probability of reaching the absorbing state is:
\begin{equation}
P_d(t,L) \sim t^{-\delta-1} {\cal{F}}(t L^{-z})
\label{survival}
\end{equation}
where ${\cal{F}}(t L^{-z})$ is an unspecified scaling function, controlling the finite-size effects,
 $z$ is the dynamical exponent and $L$ is --in general-- a measure of the linear size of the system \cite{munoz1999avalanches}.  The case of a lattice of size $N$ we get $N\sim L^d$, where $d$ is the dimension. Thus, in general, the average lifetime is $\tau_{life}=\langle t \rangle \sim \int dt \, t^{-\delta -1} \, t \, {\cal{F}}(t N^{-z/d}) \sim {N}^{z(1-\delta)/d}$.  Since in the mean-field case $\delta=1$,  the integral results into a logarithm and one expects $\tau_{life}$ to grow logarithmically with network size $ \tau_{life} \sim \log(N)$. 
Similarly, as $\tau_{life}$ is peaked around a central 
value, the position of the peak, $\tau_{life}^{max}$, scales in the same way.

\begin{figure}
\centering
\includegraphics[width=0.49\textwidth]{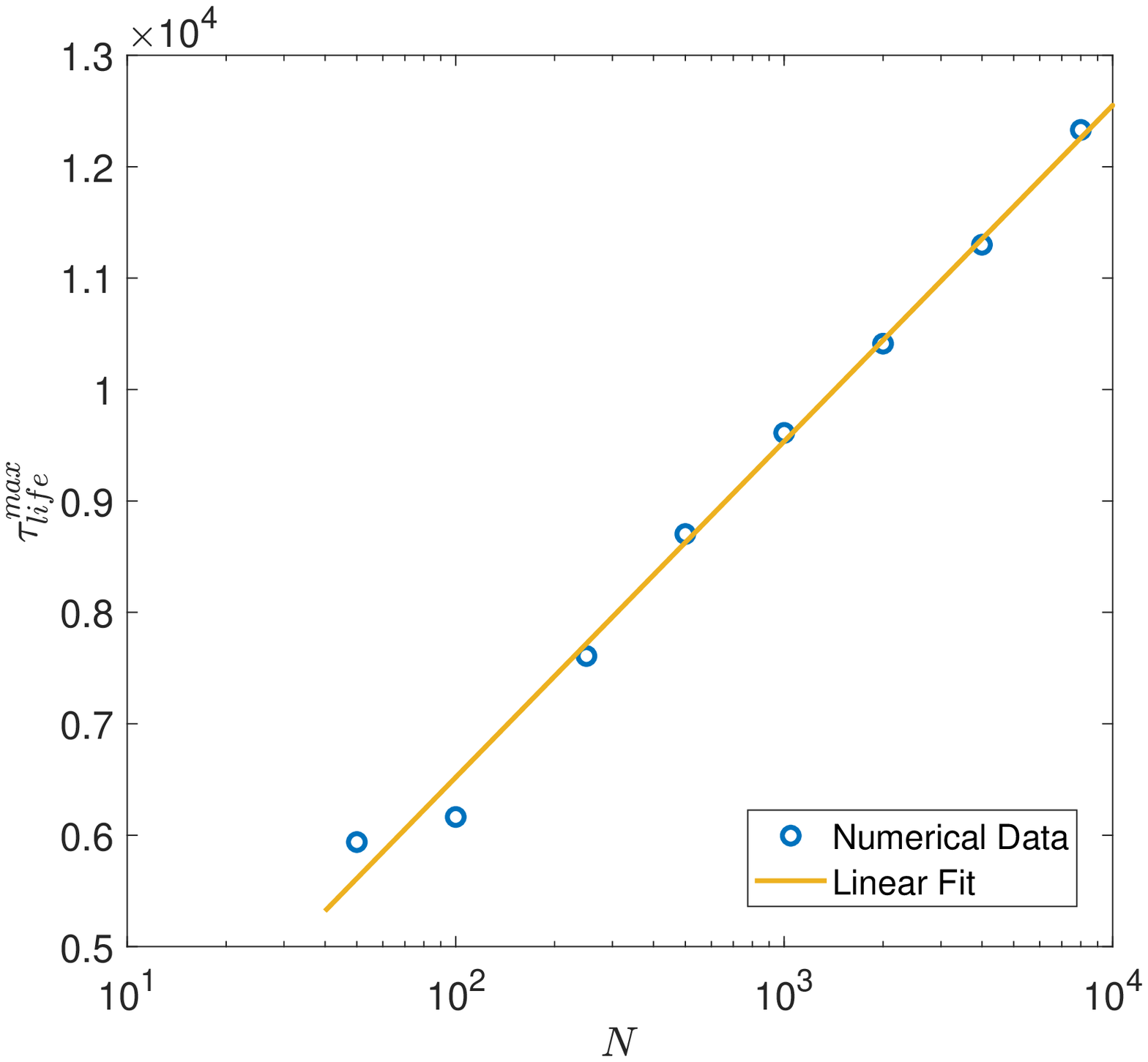}
\includegraphics[width=0.49\textwidth]{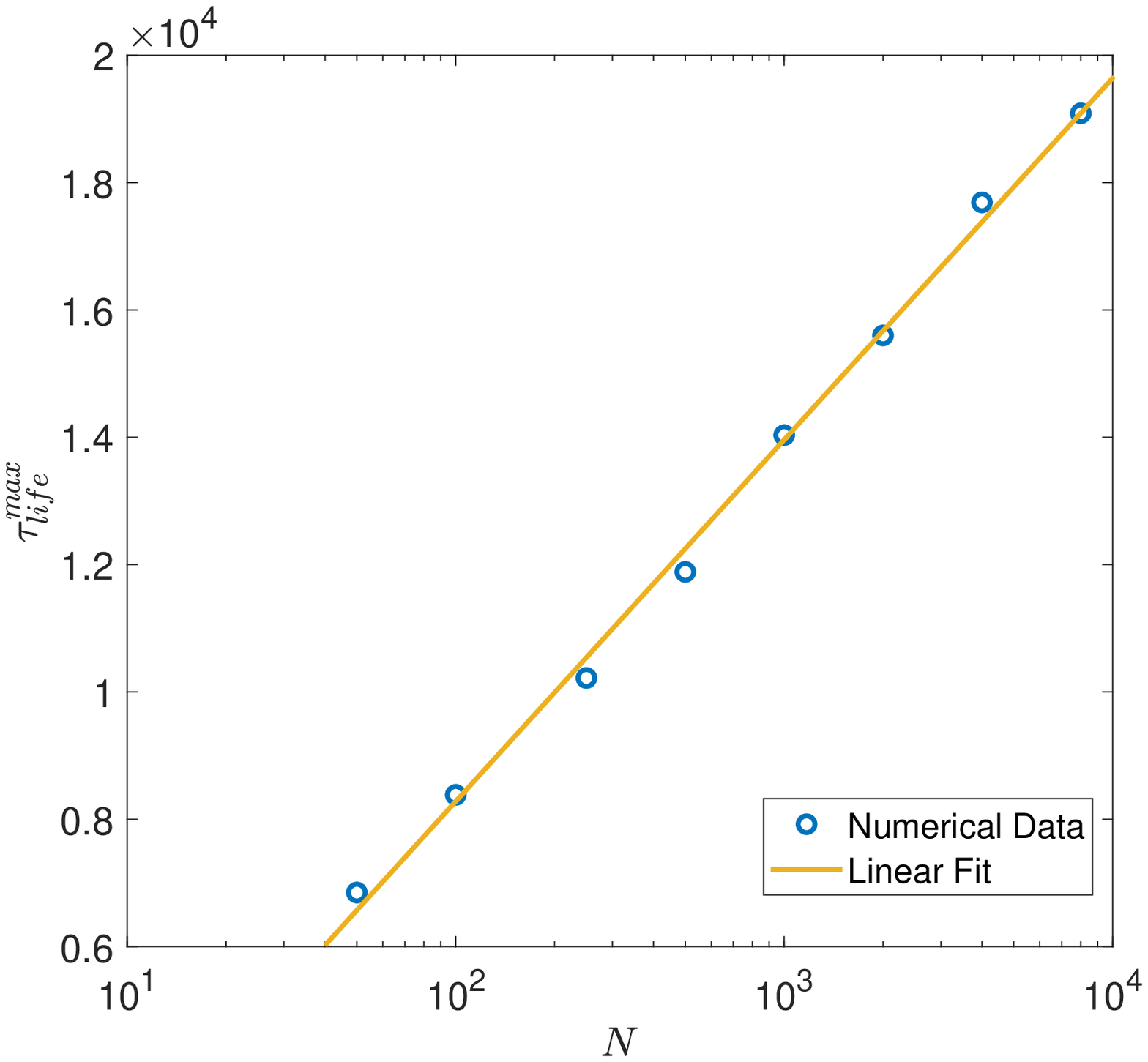}
\caption{Finite-size scaling of the lifetime: plot of $\tau_{life}^{max}$ as a function of $N$. In the left panel we consider $\Psi_a(\tau) \sim e^{-a\tau}$ (Poisson dynamics) and $\rho(a) \sim \delta(a-a_m)$, $\lambda=1$ and $a_m=10^{-4}$. In the right panel we set $\Psi_a(\tau) \sim \tau^{-(\alpha+1)}$ (bursty dynamics), $\alpha=5$ and $\rho(a) \sim \delta(a-a_m)$, $\lambda=1$ and $a_m=10^{-4}$. Blue dots represent numerical data and yellow lines represent the linear fit.}
\label{fig:Scaling}
\end{figure}

As shown in Figure \ref{fig:Scaling} this is, indeed, the case in numerical simulations for either exponential (Poisson) or power-law distributed inter-event time distributions keeping $\rho(a) \sim \delta(a-a_m)$ (see figure caption).

\begin{figure} 
\includegraphics[width=0.49\textwidth]{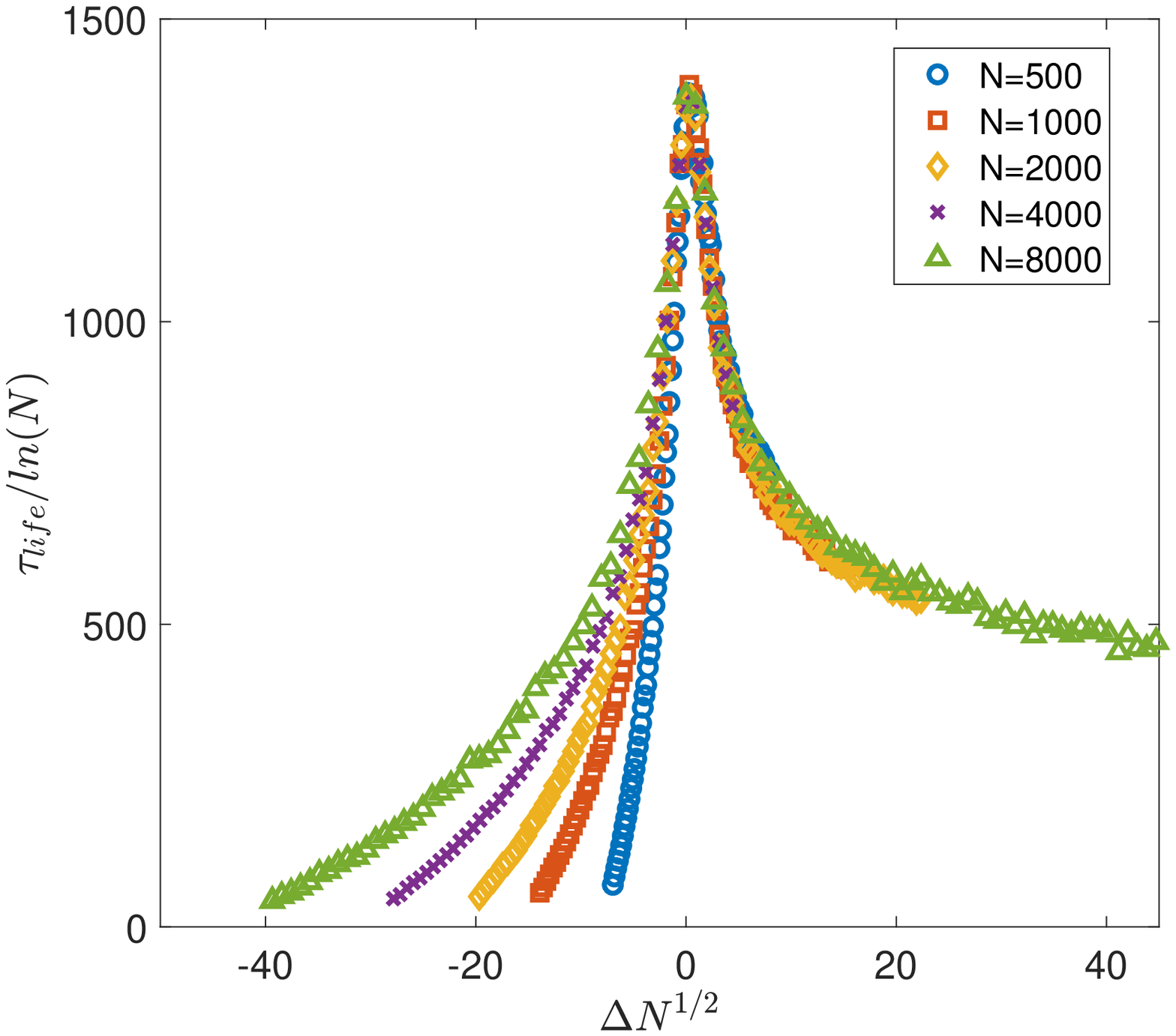} \includegraphics[width=0.49\textwidth]{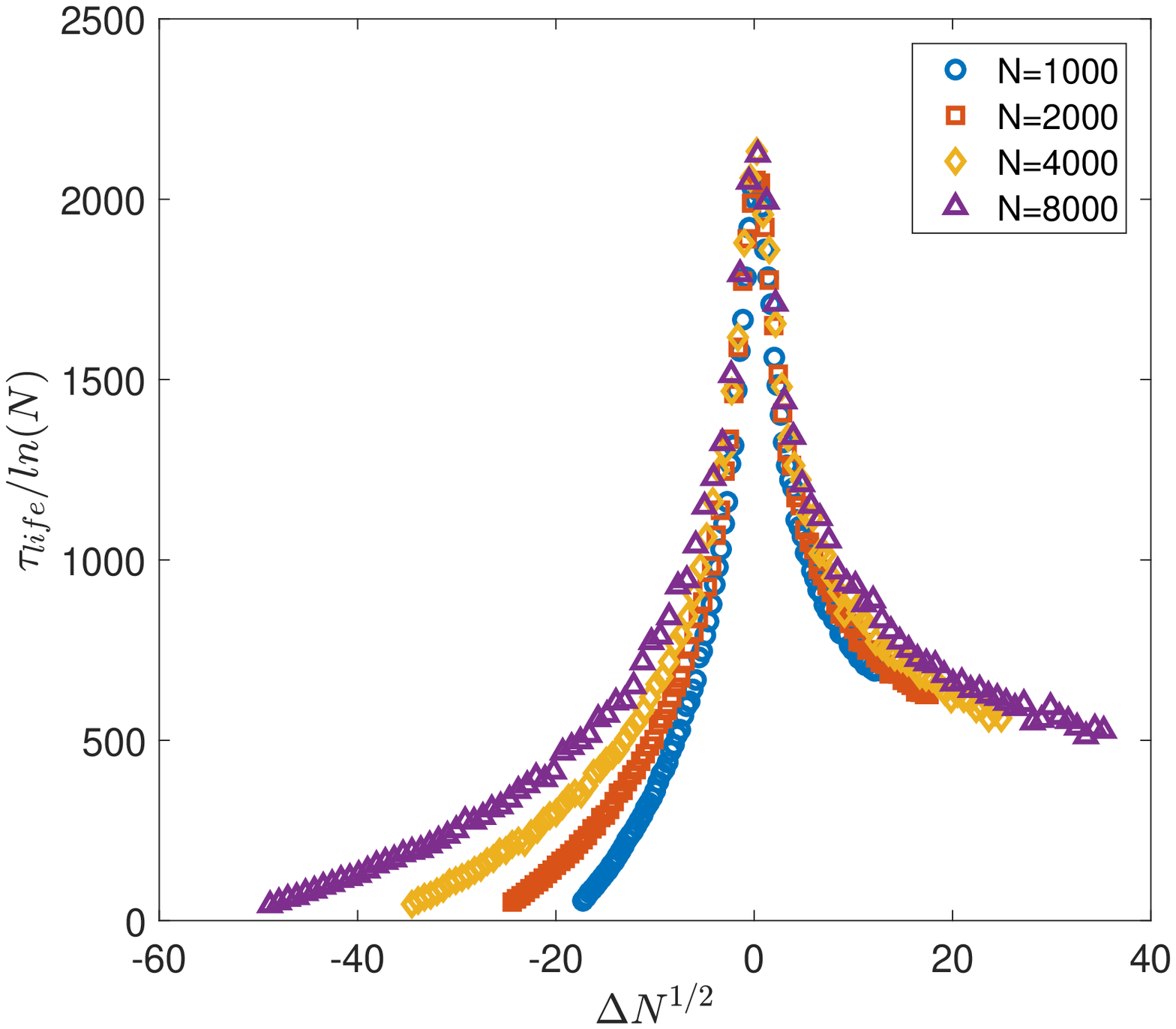} 
\caption{Rescaled lifetime: plot of the scaling function $ \mathcal{G} \left( \Delta N^{\gamma} \right)=\tau_{life}(\Delta,N)/ln(N) $ as a function of $\Delta N^{\gamma}$ fixing the exponents obtained with the finite-size scaling method. In the left panel we consider $\Psi_a(\tau) \sim e^{-a \tau}$ (Poisson process), $\rho(a) \sim \delta(a-a_m)$, $\lambda =1$, $a_m = 1.65 \, 10^{-4}$ and the exponents $ \delta = 1 $ and $\gamma= 1/2$. In the right panel we consider $\Psi_a(\tau) \sim \tau^{-(\alpha+1)}$ (bursty dynamics) with $\alpha=5$, $\rho(a) \sim \delta(a-a_m)$, $\lambda =1$, $a_m =1.65 \, 10^{-4}$ and the exponents $ \delta = 1 $ and $\gamma= 1/2$.  } 
\label{fig:Riscalato} 
\end{figure}

In the light of the above, let us adopt a finite-size scaling approach -- similar to that proposed in \cite{mata2015lifespan} -- to analyze lifespans (such as those shown in Figure \ref{fig:Numerico}) as a function of $N$ and the distance $\Delta$ to the critical point: $\Delta \equiv(r-r_c)/r_c$.  Since right at the critical point ($\Delta=0$) lifespans grows as $\ln(N)$, slightly away from the epidemic threshold we get the finite-size scaling relation:
\begin{equation}
\tau_{life}(\Delta,N) \sim \ln(N) {\cal{G}}(\Delta N^\gamma)
\label{eq:scalinghp}
\end{equation}  
where $\gamma$ is expected to take its mean-field value $1/2$ \footnote{Actually, the characteristic length-scale scales in mean field as $L \sim \Delta^{-1/2}$ and $N \sim L^{d}$, where in mean-field $d=d_C=4$ is the critical dimension, so that $N\sim \Delta^{-2}$ and, thus $\gamma=1/2$ \cite{munoz1999avalanches}.}  Finally, in Figure \ref{fig:Riscalato} we verify the scaling hypothesis of Eq. \eqref{eq:scalinghp} both in the exponential and in the power law case, showing a rather good curve collapse when rescaling (dividing) $\tau_{life}$ with $\ln(N)$ and $\Delta$ with a factor $N^{\gamma=1/2}$.  

\begin{figure}
\centering
\includegraphics[width=0.5\textwidth]{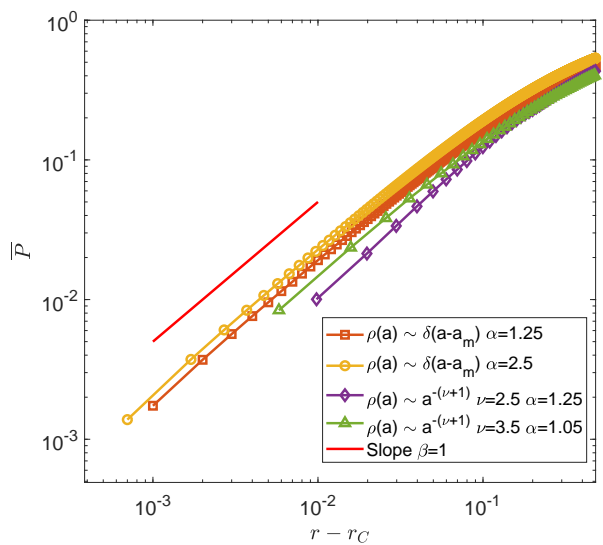}
\caption{Scaling of the prevalence: plot of $\overline{P}$ as a function of the control parameter $r-r_C$ for several values of the $\alpha$ exponent and for several activity distributions $\rho(a)$ (legend). The curves are compared with the expected slope $\beta=1$ for the mean-field type of scaling. We fixed $\lambda=1$ and the activity lower cut-off $a_m=10^{-4}$.}
\label{fig:Scaling_PO_bursty}
\end{figure}

We also measure the  critical exponent $\beta$, that describes the vanishing of the prevalence $\overline{P}$ at the epidemic threshold:
\begin{equation}
\overline{P} \sim (r-r_C)^{\beta}
\end{equation}
for $r>r_C$. In \ref{sez:fluctuation} we show that $ \beta = 1 $ for the Poisson process, regardless of $ \rho(a) $. Figure \ref{fig:Scaling_PO_bursty} shows that also in the bursty case $ \beta \simeq 1 $ for every $\alpha$ and activity distribution $\rho(a)$.

Summarizing, despite the non-trivial temporal dynamics, critical exponents suggest that burstiness does not change the universality class of the absorbing-active phase transition for the SIS model on activity-driven network, when the average inter-event time is finite ($\alpha > 1$). The system presents the same mean-field like scaling properties for non-Poissonian (bursty) and Poissonian dynamics, regardless of the exponent $\alpha$, and shows the mean-field exponents of DP universality class. A similar equivalence is obtained in \cite{starnini2017equivalence}, where burstiness is introduced on static networks as a non-Poissonian effect in infection process.

\section{Conclusions}\label{sez:conclusions}

Our results provide a coherent understanding of the interplay between the bursty dynamics of an activity-driven network and the epidemic process unfolding on it.
Focusing on the SIS model, by means of an activity-based mean-field approach we analytically derive a closed form for the epidemic threshold on an activity-driven network: the threshold depends on the whole functional form of the activity and inter-event time distributions. The mean-field approach turns out to be exact in this system, as connections are continuously reshuffled destroying local correlations. We show that the analytical results are in excellent agreement with extensive numerical simulations.

The model captures two key aspects driving the evolution of social
networks: burstiness and heterogeneity in human activities. As is well known, heterogeneity produces a decrease in the epidemic threshold.
We show that burstiness also reduces the epidemic threshold, while the effect on prevalence is more subtle: in low-infective systems burstiness raises the prevalence, while it weakens the epidemics in high infective systems. We showed that this behavior can be attributed to fluctuations effects. Finally, despite the non-trivial temporal dynamics,  the values of the scaling exponents suggest that burstiness does not change the universality class of the absorbing transition.

Our results can help to clarify the effect of burstiness on dynamical process in temporal network and, in turn, to design efficient immunization strategies \cite{lee2012exploiting} and control protocols, based on temporal structure of the human interactions. The model can be extended to different epidemic processes and  the analytic results for the thresholds should hold also in the case of the SIR model,  while a different behavior is expected in the active phase \cite{tizzani2018epidemic}. We can also include in the model further features of realistic social networks, i.e. the presence of communities \cite{nadini2018epidemic}, adaptive behavior \cite{funk2009spread,rizzo2014effect} and memory \cite{tizzani2018epidemic}, these combined with burstiness can produce non-trivial effects \cite{ubaldi2017burstiness}. 

\vspace{1cm}
\appendix
\section{Evaluation of the epidemic threshold}\label{sez:Conti}
Let us consider the system of equations \eqref{eq:system}.
We set an initial condition, adding $\varepsilon_a \delta(t)$ to $Q_a(t)$: i.e. we impose that at $t=0$ all nodes are active and a node with activity $a$ is infected with probability $\varepsilon_a$.  For  $\widetilde{Q}_a(s)$ and $\widetilde{P}_a(s)$, the Laplace transforms of $Q_a(t)$ and $P_a(t)$, we obtain:

\begin{equation}
\begin{split}
\widetilde{Q}_a(s) = \int_0^{\infty} dt \, e^{-st} \, Q_a(t) =& \frac{1}{s} \left[ \frac{\lambda \overline{aQ}}{\mu + \lambda \overline{aQ}} + \left( \lambda \overline{P} - \frac{\lambda \overline{aQ}}{\mu + \lambda \overline{aQ}} \right) L_a(\mu+\lambda \overline{aQ}) \right]\\
&+ (1-\lambda \overline{P}) \widetilde{Q}_a(s) \int_0^{\infty} d\tau' e^{-(\mu+\lambda \overline{aQ}+s)\tau'} \Psi_a(\tau') + \varepsilon_a
\end{split}
\end{equation}
\begin{equation}
\begin{split}
\widetilde{P}_a(s) = \int_0^{\infty} dt \, e^{-st} \, P_a(t) =& \frac{1}{s} \left[ \frac{\lambda \overline{aQ}}{\mu + \lambda \overline{aQ}} + a \left( \lambda \overline{P}- \frac{\lambda \overline{aQ}}{\mu + \lambda \overline{aQ}} \right) \frac{1-L_a(\mu+ \lambda \overline{aQ})}{\mu + \lambda \overline{aQ}} \right]\\
&+ a (1-\lambda \overline{P}) \widetilde{Q}_a(s) \int_0^{\infty} d\tau' e^{-(\mu + \lambda \overline{aQ} +s )\tau'} \int_{\tau'}^{\infty} d\tau'' \Psi_a(\tau'')
\end{split}
\end{equation}
We substitute $\widetilde{Q}_a(s)$ in $\widetilde{P}_a(s)$ and expand both for $s \sim 0$, since for the calculation of the epidemic threshold we only need the asymptotic behavior ($t \rightarrow \infty$). We obtain at first order in $\mathcal{O}(\frac{1}{s})$:

\begin{equation}
\widetilde{Q}_a(s) = \frac{C_1}{s} + \mathcal{O}(\frac{1}{s})
\label{eq:Q_tilde}
\end{equation}
\begin{equation}
\widetilde{P}_a(s) = \frac{C_3}{s} + \frac{C_1}{s} a (1-\lambda \overline{P})  \frac{1-L_a(\mu+ \lambda \overline{aQ})}{\mu + \lambda \overline{aQ}} + \mathcal{O}(\frac{1}{s})
\end{equation}

where
$$C_1= \frac{\left[ \frac{\lambda \overline{aQ}}{\mu + \lambda \overline{aQ}} + \left( \lambda \overline{P} - \frac{\lambda \overline{aQ}}{\mu + \lambda \overline{aQ}} \right)L_a(\mu+ \lambda \overline{aQ}) \right]}{1-(1-\lambda \overline{P})L_a(\mu+ \lambda \overline{aQ})} \,\,\, , \,\,\,  L_a(s)= \int_0^{\infty} d\tau \, e^{-s \tau} \, \Psi_a(\tau)$$

$$C_3 = \frac{\lambda \overline{aQ}}{\mu+\lambda \overline{aQ}} + a \left( \lambda \overline{P} - \frac{\lambda \overline{aQ}}{\mu+\lambda \overline{aQ}} \right) \frac{1-L_a(\mu+\lambda \overline{aQ})}{\mu + \lambda \overline{aQ}}$$
Going back to the temporal domain, in the asymptotic limit we obtain:
\begin{equation*}
Q_a(t) \xrightarrow[t \rightarrow \infty]{} C_1=Q_a^0 \,\,\, , \,\,\, 
P_a(t) \xrightarrow[t \rightarrow \infty]{} C_4=P_a^0 
\end{equation*}
where\\
$$C_4=C_3 + C_1 a(1-\lambda \overline{P}) \frac{1-L_a(\mu+ \lambda \overline{aQ})}{\mu+\lambda \overline{aQ}} = C_3 + a(1-\lambda \overline{P}) \frac{1-L_a(\mu+ \lambda \overline{aQ})}{\mu+\lambda \overline{aQ}} Q_a^0$$
In this way we obtain two self-consistency equations for $Q_a^0$ and $P_a^0$:
\begin{equation}
P_a^0=\frac{\lambda \overline{aQ}}{\mu + \lambda \overline{aQ}} + a \frac{1-L_a(\mu+ \lambda \overline{aQ})}{(\mu + \lambda \overline{aQ})^2} \frac{\lambda \mu \overline{P}}{1-(1-\lambda \overline{P})L_a(\mu+ \lambda \overline{aQ})}
\label{eq:Pa0}
\end{equation}
\begin{equation}
Q_a^0=\frac{\left[ \frac{\lambda \overline{aQ}}{\mu + \lambda \overline{aQ}} + \left( \lambda \overline{P} - \frac{\lambda \overline{aQ}}{\mu + \lambda \overline{aQ}} \right)L_a(\mu+ \lambda \overline{aQ}) \right]}{1-(1-\lambda \overline{P})L_a(\mu+ \lambda \overline{aQ})}
\label{eq:Qa0}
\end{equation}
since $\overline{P} = \int_0^{\infty} da \, \rho(a) \, P_a^0$ and $\overline{aQ}= \int_0^{\infty} da \, a \, \rho(a) \, Q_a^0$. Integrating the above expression over $a$ we get the self consistence equations described in Eq. \eqref{eq:mappa}.

\section{Prevalence at $r=1/\lambda$}\label{sez:prevalence}

We can write Eqs. \eqref{eq:Pa0} and \eqref{eq:Qa0} for $P_a^0$ and $Q_a^0$ as:
\begin{equation}
P_a^0(\mu+\lambda \overline{aQ})^2 = \lambda \overline{aQ}(\mu+\lambda \overline{aQ}) + \lambda \mu \overline{P} \frac{a (1-L_a(\mu+ \lambda \overline{aQ}))}{1-(1-\lambda \overline{P})L_a(\mu+ \lambda \overline{aQ})}
\label{eq-prev1}
\end{equation}
\begin{eqnarray}
Q_a^0 (\mu+\lambda \overline{aQ}) =  \lambda \overline{aQ} +  \frac{\mu\lambda \overline{P} L_a(\mu+ \lambda \overline{aQ})}{1-(1-\lambda \overline{P})L_a(\mu+ \lambda \overline{aQ})}
\label{eq-prev2}
\end{eqnarray}
We multiply Eq. \eqref{eq-prev1} by $\rho(a)$ and Eq. \eqref{eq-prev2} by $\rho(a)a/\mu$. Then we integrate over the activity obtaining for $\overline{aQ}$ and $\overline{P}$:
\begin{equation}
-\mu \overline{P}+\lambda B \overline{P} - \frac{\lambda^2}{\mu} \overline{aQ}^2 (\overline{P}-1) + \lambda \overline{aQ}-2\lambda \overline{aQ}\, \overline{P}=0
\label{eq:P_burst}
\end{equation}
\begin{equation}
\frac{\lambda}{\mu} \overline{aQ}^2 + \left( 1-\frac{\lambda \langle a \rangle}{\mu} \right) \overline{aQ}-\lambda H \overline{P}=0
\label{eq:aQ_burst}
\end{equation}
where $B= \int_0^{\infty} da \, \rho(a) \frac{a(1-L_a(\mu+ \lambda \overline{aQ}))}{1-(1-\lambda\overline{P})L_a(\mu+ \lambda \overline{aQ})}$, $H= \int_0^{\infty} da \, \rho(a) \frac{aL_a(\mu+ \lambda \overline{aQ})}{1-(1-\lambda \overline{P})L_a(\mu+ \lambda \overline{aQ})}$.
Setting $\langle a \rangle / \mu =r = 1/\lambda$, Eq. \eqref{eq:aQ_burst} gives:
\begin{equation}
\overline{aQ}^2 / \mu=  H \overline{P}
\label{eq:aQ_quad}
\end{equation}
Plugging Eq. \eqref{eq:aQ_quad} in Eq. \eqref{eq:P_burst} and setting $\mu=\lambda \langle a \rangle$, we obtain:
\begin{equation}
\overline{P}(B-\langle a \rangle - \lambda H(\overline{P}-1))+\overline{aQ}(1-2\overline{P})=0
\label{eq-prev3}
\end{equation}
Since
\begin{eqnarray}
B-\langle a \rangle - \lambda H (\overline{P}-1) &=& \int_0^{\infty} da \rho(a) \left[ \frac{a(1-L_a(\mu+ \lambda \overline{aQ}))}{1-(1-\lambda \overline{P})L_a(\mu+ \lambda \overline{aQ})} \right. \nonumber \\
&&- a- \left. \lambda \overline{P} \frac{aL_a(\mu+ \lambda \overline{aQ})}{1-(1-\lambda \overline{P})L_a(\mu+ \lambda \overline{aQ})} \right] + \lambda H \nonumber \\
&=& \lambda H (1-2\overline{P})
\end{eqnarray}
then Eq. \eqref{eq-prev3} becomes:
\begin{equation}
(1-2\overline{P})(\overline{aQ}+\lambda H \overline{P})=0
\label{eq:P_bursty_1_2}
\end{equation}
Since $H>0$, $\overline{aQ}\geq0$ and $\overline{P}\geq0$, Eq.  \eqref{eq:P_bursty_1_2} has only two solutions: the absorbing state $(\overline{P},\overline{aQ})=(0,0)$ which is unstable because from Eq. \eqref{eq:disug_generale} we get $r_C< 1/\lambda=r$; and $\overline{P}=1/2$ which represents an endemic state, stable at $r= 1/\lambda$.

In this way we proved that $\overline{P}=1/2$ if $r=1/\lambda$ regardless of the inter-event time distribution $\Psi_a(\tau)$ and of the activity distribution $\rho(a)$.
Therefore, as observed in Fig. \ref{fig:Prevalence}, the prevalences $\overline{P}$ calculated for different $\alpha$ cross at $r=1/\lambda$ and so there is a switch at $r=1/\lambda$ in the effect of burstiness on epidemic spreading. 

\section{Exponential inter-event time distributions}\label{sez:fluctuation}

In this section we discuss two properties of the model with exponential inter-event time distribution.  First we calculate exactly the critical exponent $\beta=1$. 
In this case the probability $P_a(t)$ is governed by the standard activity based mean-field differential equation:
\begin{equation}
\partial_t P_a(t) = -\mu P_a(t) + \lambda (1-P_a(t)) [a \overline{P}(t)+\overline{aP}(t)]
\label{eq:Pa_beta}
\end{equation}
Imposing the steady state condition $\partial_t P_a(t)=0$ we get the stationary value $P_a^0$ of the $P_a(t)$:
\begin{equation}
0 = -\mu P_a^0 + \lambda (1-P_a^0)(a \overline{P}+ \overline{aP})
\label{eq:Pa_aP}
\end{equation}
Averaging Eq. \eqref{eq:Pa_aP}  over $a$ we get $\overline{aP}$ as a function of $\overline{P}$:
\begin{equation}
\overline{aP}=\frac{\mu (\lambda r -1)}{\lambda (2\overline{P}-1)} \overline{P}
\label{eq:aP_beta}
\end{equation}
Plugging Eq. \eqref{eq:aP_beta} in Eq. \eqref{eq:Pa_aP} we obtain for $P_a^0$:
\begin{equation}
P_a^0 = \frac{\lambda a \overline{P} (2 \overline{P}-1)+ \mu (\lambda r -1) \overline{P}}{\lambda a \overline{P} (2 \overline{P}-1)+ \mu (\lambda r -1) \overline{P} + \mu (2\overline{P}-1)}
\label{eq:Pa0_beta}
\end{equation}
Multiplying both members for $\rho(a)$ and integrating over the activity we obtain a self-consistency equation for $\overline{P}$:
\begin{equation}
\overline{P}= \overline{P} \int_0^{\infty} da \rho(a) \frac{ \frac{2 \lambda a}{\mu} \overline{P} + \lambda r -1 - \frac{\lambda a}{\mu}}{ \frac{2 \lambda a}{\mu} \overline{P}^2 +  (\lambda r +1 - \frac{\lambda a}{\mu})\overline{P} -1}
\label{eq:P_beta_1}
\end{equation}
Eq. \eqref{eq:P_beta_1} admits the solution $\overline{P}=0$ representing the absorbing state. Other possible solutions can be obtained by solving for $\overline{P}$  the equation:
\begin{equation}
1 = \int_0^{\infty} da \rho(a) \frac{ \frac{2 \lambda a}{\mu} \overline{P} + \lambda r -1 - \frac{\lambda a}{\mu}}{ \frac{2 \lambda a}{\mu} \overline{P}^2 +  (\lambda r +1 - \frac{\lambda a}{\mu})\overline{P} -1}
\label{eq:Expand1}
\end{equation}
To evaluate $\beta$ we assume $r \sim r_C$ and we can expand Eq. \eqref{eq:Expand1} near the critical point,  for $\overline{P} \sim 0$. We get:
\begin{eqnarray}
1 = \int_0^{\infty} & & da \rho(a) \left[ \frac{2 \lambda a}{\mu} \overline{P} + \lambda r -1 - \frac{\lambda a}{\mu} \right] \left[ -1+ \overline{P} \left( \frac{\lambda a }{\mu} - \lambda r -1 \right) \right. \nonumber\\
& & \left.
+ \overline{P}^2 \left( \frac{2 \lambda^2 a }{\mu} r - \lambda^2 r^2 -1 -\frac{\lambda^2 a^2}{\mu^2}-2\lambda r \right) + \mathcal{O}(\overline{P}^2) \right]
\label{eq:Expand2}
\end{eqnarray}

Integrating over the activity and keeping only the leading orders in $\overline{P}$ we get:
\begin{equation}
\overline{P} \left[ \overline{P} C(r) -  (r-r_C)D(r) \right]=0
\label{eq:Expand3}
\end{equation}
where 
\begin{equation}
C(r)=1+ \frac{\lambda^3 \langle a^3 \rangle}{\mu^3} + 2 \lambda^3 r^3 - 3 \lambda^2 r^2 + 3 \frac{\lambda^2 \langle a^2 \rangle}{\mu^2} (1-\lambda r)
\end{equation}
 and 
\begin{equation}
D(r)=\lambda^2\left( \frac{ \langle a^2 \rangle}{\lambda^2 \langle a \rangle^2} -1 \right)\left( \frac{ \langle a \rangle}{\lambda (\sqrt{\langle a^2 \rangle}-\langle a \rangle )} +r \right)
\end{equation}
and for the exponential case
\begin{equation}
r_C=  \frac{ \langle a \rangle}{\lambda (\sqrt{\langle a^2 \rangle}+\langle a \rangle )}
\label{thresh_pois}
\end{equation}

Since $\langle a^2\rangle>\langle a \rangle^2$ we have that $D(r)>0$. Moreover close to the critical point:
$$C(r_C) = \frac{\langle a^3 \rangle + 4 \langle a^2 \rangle ^{3/2} + 3 \langle a \rangle \langle a^2 \rangle}{(\langle a \rangle + \sqrt{\langle a^2 \rangle})^3}>0$$

\begin{figure}
\centering
\includegraphics[width=0.5\textwidth]{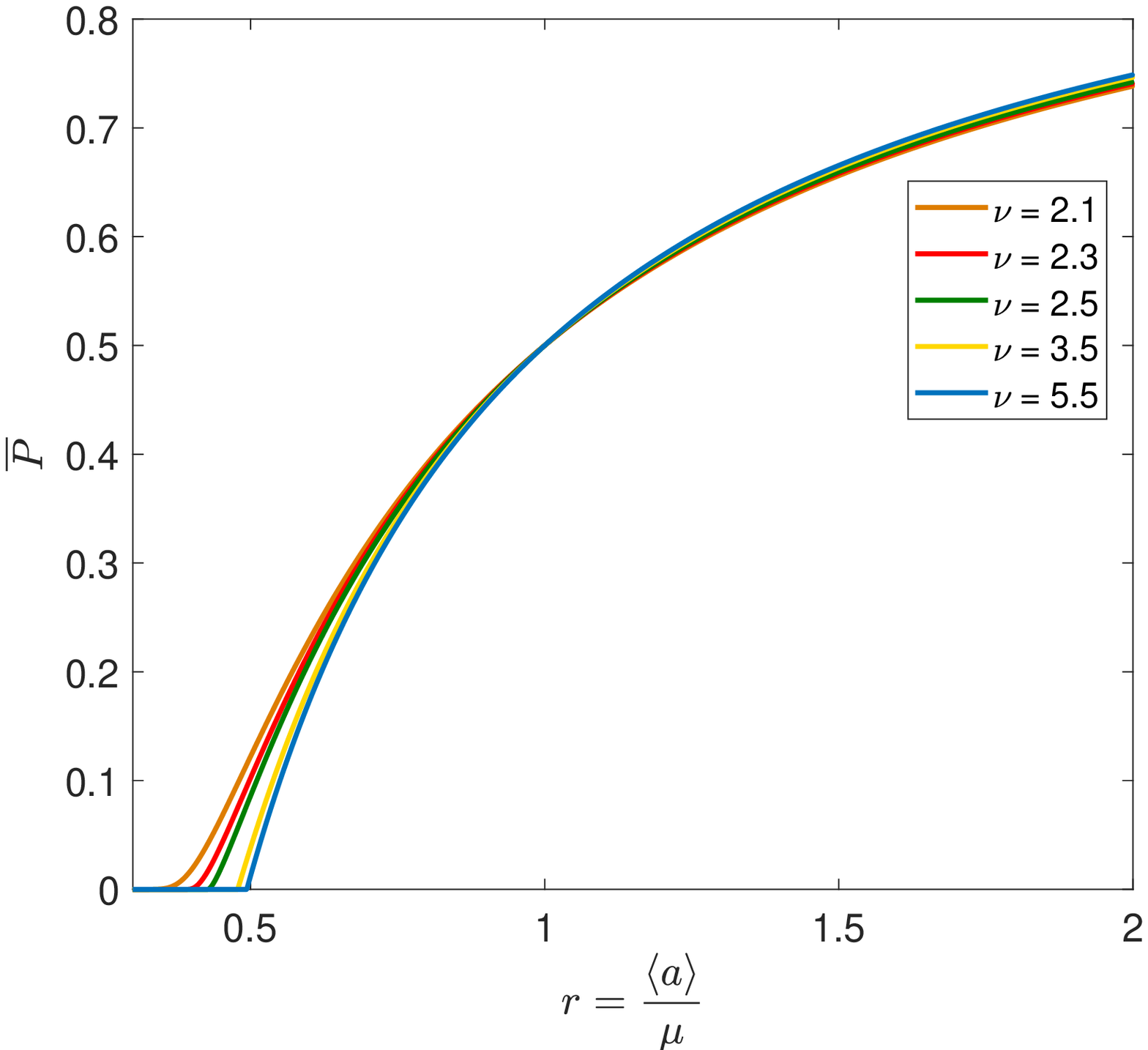}
\caption{Prevalence $\overline{P}$ as a function of the control parameter $r= \langle a \rangle / \mu$ for $\Psi_a(\tau) \sim e^{-a\tau}$ and heterogeneously distributed activities $\rho(a) \sim a^{-(\nu+1)}$. We fix the lower cut-off $a_m=10^{-4}$, $\lambda=1$ and consider several values of $\nu \in [2.1,5.5]$.}
\label{fig:fluctuation_activity}
\end{figure}

Hence for $r\sim r_C$ Eq. \eqref{eq:Expand3} has two solutions: $\overline{P}=0$ i.e. the absorbing state, and for $r>r_C$ 
\begin{equation}
\overline{P}(r)=(r-r_C)\frac{D(r_C)}{C(r_C)} 
\label{eq:Expand4}
\end{equation}
which means that $\beta=1$.

We now consider the behavior of the prevalence $\overline{P}(r)$ in the active phase as a function of the activity fluctuations, i.e. we vary the exponent $\nu$ characterizing the distribution $\rho(a)$ defined by Eq. \eqref{activity}.  We notice that, according to Eq. \eqref{thresh_pois},  the threshold $r_C$ increases by increasing $\nu$, hence at small $\overline{P}$ close to the epidemic thresholds the prevalence grows with $\nu$. Then the results of \ref{sez:prevalence} show that at $r=1/\lambda> r_C$, for any values of $\nu$, $\overline{P}=1/2$,  i.e. the prevalence is independent of activity fluctuations. Finally, for $r>1/\lambda$ the opposite behavior is expected i.e.  the system presents a larger prevalence at large $\nu$. The overall behavior is illustrated in Figure \ref{fig:fluctuation_activity}: large activity fluctuations and large burstiness give rise to a similar effect on the prevalence i.e. an increase at small $r$ and a decrease at large $r$.

\section*{Acknowledgements}
R.B. acknowledges funding from the INFN BIOPHYS project. M.A.M. acknowledges the Spanish Ministry of Science as well as the Agencia Espa{\~n}ola de Investigaci{\'o}n (AEI) for financial support under grant FIS2017-84256-P (FEDER funds). We warmly thank V. Buendia for very useful comments.

\section*{References}
%\bibliography{Biblio_paper}
\providecommand{\newblock}{}

\end{document}